\documentclass{article}

\usepackage{PRIMEarxiv}

\usepackage[utf8]{inputenc} 
\usepackage[T1]{fontenc}    
\usepackage{hyperref}       
\usepackage{url}            
\usepackage{booktabs}       
\usepackage{amsfonts}       
\usepackage{nicefrac}       
\usepackage{microtype}      
\usepackage{lipsum}
\usepackage{fancyhdr}       
\usepackage{graphicx}       
\usepackage{tikz}           
\usepackage{subcaption}     
\usepackage[dvipsnames]{xcolor}  
\graphicspath{{media/}}     

\usepackage{amssymb}
\usepackage{amsmath}
\usepackage{amsthm}
\usepackage{xcolor}
\usepackage{natbib}
\usepackage{standalone}
\usepackage{cleveref}

\title{Stationary subspace analysis for spatial data
}

\author{
  Perttu Saarela \\
  Department of Mathematics and Statistics\\ University of Helsinki\\ 
  Finland\\
  \texttt{perttu.saarela@helsinki.fi} \\
   \And
   Klaus Nordhausen \\
  Department of Mathematics and Statistics\\ University of Helsinki\\ 
  Finland\\
  \texttt{klaus.nordhausen@helsinki.fi} \\
   \AND
  Jaakko Pere \\
   School of Business\\  Aalto University\\ 
  Finland\\
  \texttt{jaakko.pere@aalto.fi} \\
   \And
  Anne M. Ruiz \\
  Toulouse School of Economics\\ University Toulouse Capitole\\ 
  France\\
  \texttt{anne.ruiz-gazen@tse-fr.eu} \\
}

\DeclareMathOperator*{\argmin}{arg\,min}

\begin{document}
\maketitle

\providecommand{\mlt}{\mathrm{MLC}}
\providecommand{\mcd}{\mathrm{MCD}}
\providecommand{\diag}{\mathrm{diag}}
\providecommand{\off}{\mathrm{off}}
\providecommand{\med}{\mathrm{med}}
\providecommand{\MAD}{\mathrm{MAD}}
\newtheorem{theorem}{Theorem}[section]
\newtheorem{corollary}{Corollary}[theorem]
\newtheorem{lemma}[theorem]{Lemma}

\theoremstyle{definition}
\newtheorem{remark}{Remark}

\newcommand{\comm}[1]{\textcolor{purple}{#1}}
\newcommand{\indep}{\bot\!\!\,\!\!\bot}

\newcommand{\bs}[1]{\boldsymbol{#1}}
\providecommand{\bo}[1]{\mathbf{#1}}
\newcommand{\x}{\bo x}
\newcommand{\xu}{\x(\bo u)}
\newcommand{\xst}{\x^{st}}
\newcommand{\xstu}{\x^{st}(\bo u)}
\newcommand{\z}{\bo z(\bo u)}
\newcommand{\n}{\bo n(\bo u)}
\newcommand{\s}{\bo s(\bo u)}
\newcommand{\e}{\bs\varepsilon}
\newcommand{\h}{\bo h}
\newcommand{\m}{\bo m}
\newcommand{\A}{\bo A}
\newcommand{\B}{\bo B}
\newcommand{\C}{\bo C}
\newcommand{\W}{\bo W}
\newcommand{\D}{\bo D}
\newcommand{\U}{\bo U}
\renewcommand{\u}{\bo u}

\newcommand{\V}{\bo V}
\newcommand{\T}{\bo T}
\newcommand{\M}{\bo M}
\newcommand{\Cov}{\mathrm{Cov}}
\newcommand{\Lcov}{\mathrm{Lcov}}
\newcommand{\var}{\mathrm{var}}
\newcommand{\mean}{\mathrm{mean}}
\newcommand{\cor}{\mathrm{cor}}
\newcommand{\comb}{\mathrm{comb}}
\newcommand{\E}{\mathrm{E}}
\newcommand{\dom}{\mathcal{U}}
\newcommand{\R}{\mathbb{R}}
\newcommand{\aug}{\mathrm{AUG}}

\newcommand{\norm}[1]{\left\lVert#1\right\rVert}

\begin{abstract}
Stationary subspace analysis (SSA) is a blind source separation framework that
decomposes linearly mixed multivariate data into stationary and nonstationary
components. We extend SSA to spatially indexed data by introducing spatial
stationary subspace analysis (spSSA), which explicitly accounts for spatial
dependence. We propose three estimation procedures for the unmixing matrix based
on first- and second-order spatial statistics. Each procedure targets a
different type of nonstationarity and can be formulated as the solution to a
generalized eigenvalue problem. To address situations where multiple forms of
nonstationarity are present simultaneously, we combine the three procedures
using approximate joint diagonalization. Simulation studies demonstrate that
this combined approach yields superior separation performance. When the
dimension of the nonstationary subspace is known, the proposed methods reliably
recover the latent stationary and nonstationary components. However, determining
this dimension remains a fundamental challenge in SSA, for which no generally
accepted solution currently exists. Building on our estimation procedures, we
propose a novel data augmentation approach to estimate the dimension of the
nonstationary subspace and demonstrate its effectiveness through simulation
studies. The proposed methodology is easily transferable to time series
settings, making it of broader methodological interest.
\end{abstract}


\section{Introduction}

In spatial data analysis, observations $x(\u_i)\in \R$, $i=1,\ldots,n$, are
collected over a spatial domain $\dom \subset \R^k$, where $\u_i \in \dom$
denotes the sampling location. A central principle guiding spatial modeling is
\emph{Tobler’s First Law of Geography} \citep{Tobler1970}, which states that
``everything is related to everything else, but near things are more related
than distant things.'' Consequently, observations recorded at nearby locations
tend to be more similar than those farther apart, and this spatial dependence
must be explicitly accounted for.

Spatial dependence is commonly characterized through the covariance function
\begin{equation*}
 C_x(\u_i, \u_j)
 =
 \E\!\left[
   \bigl(x(\u_i)-\E[x(\u_i)]\bigr)
   \bigl(x(\u_j)-\E[x(\u_j)]\bigr)
 \right].
\end{equation*}
For more tractable modeling, spatial data are often assumed to arise from a
weakly stationary square-integrable random field, implying a constant mean
$\E[x(\u)]=\mu$ and a covariance depending only on the separation vector
$\h=\u_i-\u_j$. If the covariance depends solely on $h=\|\h\|$, the process is
further said to be isotropic. Under these assumptions, several parametric
covariance models exist, with the Mat\'ern family of functions being among the
most widely used~\citep{GuttorpGneiting2006}.

In many applications, several variables are observed at each spatial location,
resulting in multivariate spatial data. Let $\xu$ denote a $p$-variate random
field with the cross-covariance function
\begin{equation*}
 \C_{\x}(\u_i, \u_j)
 =
 \E\!\left[
   (\x(\u_i)-\E[\x(\u_i)])
   (\x(\u_j)-\E[\x(\u_j)])^\top
 \right].
\end{equation*}
Constructing flexible and valid multivariate covariance models is substantially
more challenging than in the univariate case, as reviewed in
\citet{GentonKleiber2015}. A widely used framework is the linear model of
coregionalization (LMC) \citep{GoulardVoltz1992,Wackernagel2003}, which
represents the cross-covariance as a finite sum of coregionalization matrices
multiplied by univariate spatial correlation functions. Multivariate extensions
of the Mat\'ern covariance have been proposed by
\citet{GneitingKleiberSchlather2010} and \citet{ApanasovichGentonSun2012},
though validity requires nontrivial parameter constraints.

Despite their popularity, these models rely on assumptions of global weak
stationarity and isotropy, which are often unrealistic for large or
heterogeneous spatial domains. Stationarity may be plausible locally, but it
rarely holds globally. This has motivated extensive work on nonstationary
spatial models, particularly in the univariate setting. Prominent approaches
include spatial deformations \citep{SampsonGuttorp1992,Sampson2010}, kernel
convolution and closed-form nonstationary covariances
\citep{PaciorekSchervish2006}, spectral-domain methods \citep{Fuentes2002},
locally stationary estimation \citep{AnderesStein2011}, and SPDE-based
(Stochastic Partial Differential Equation) models with spatially varying
dependence parameters
\citep{LindgrenRueLindstrom2011,IngebrigtsenRueSteinsland2015}. 

Developing nonstationary models for multivariate spatial data is even more
demanding, as valid cross-covariances must simultaneously accommodate spatially
varying marginal and cross dependence \citep{VuZammitMangionCressie2022}.
Extensions of the LMC with spatially varying parameters
\citep{GelfandSchmidtBanerjeeSirmans2004} and nonstationary multivariate
Mat\'ern models \citep{KleiberNychka2012} have been proposed, as well as
multivariate spatial deformation approaches \citep{VuZammitMangionCressie2022}.
Nevertheless, these methods still rely on explicit specification of (possibly
nonstationary) cross-covariance structures and become increasingly impractical
as the dimension $p$ grows.

Indeed, as $p$ increases, the number of marginal and cross-covariance functions
grows quadratically, leading to severe computational and identifiability issues.
These challenges have fueled interest in \emph{dimension reduction} techniques,
which model multivariate spatial fields through a smaller number of latent
components capturing the dominant dependence structure. Besides its role as a
covariance construction, the LMC may be viewed as an implicit dimension
reduction approach. More explicit methods include \emph{spatial blind source
separation} (SBSS)
\citep{NordhausenOjaFilzmoserReimann2015,BachocGentonNordhausenRuizGazenVirta2020,MuehlmannBachocNordhausenYi2024},
which assumes that the observed field is a linear mixture of latent,
second-order uncorrelated spatial sources with distinct dependence structures.
Estimation typically relies on the joint diagonalization of the spatial
covariance matrices across multiple lags, after which the latent components can
be modeled individually. Extensions allowing for spatial nonstationarity, often
referred to as \emph{spatial nonstationary source separation} (SNSS)
\citep{MuehlmannBachocNordhausen2022,SipilaNordhausenTaskinen2024}, permit
spatially varying variance or dependence structures of the latent sources.

Despite their advantages, existing source separation methods impose stationarity
assumptions uniformly across all latent components, or require all components to
be nonstationary in a similar manner. To the best of our knowledge, the
possibility that the latent space decomposes into a stationary subspace and a
nonstationary subspace has not yet been systematically explored in the spatial
statistics literature. This stands in marked contrast to the time series
context, where \emph{stationary subspace analysis} (SSA) \citep[see
e.g.][]{BunauMeineckeKiralyMuller:2009,Haraetal:2010,FlumianMatilainenNordhausenTaskinen2024}
is a well-established and actively researched framework for decomposing
multivariate signals into stationary and nonstationary components. Such a
decomposition is nevertheless appealing in spatial applications, where
large-scale latent factors may induce nonstationarity, while smaller-scale
components remain approximately stationary.

The goal of this paper is to develop \emph{spatial stationary subspace analysis}
(spSSA), a novel dimension reduction framework for multivariate spatial data
that explicitly separates stationary and nonstationary latent subspaces. The
proposed methodology allows for the detection of nonstationarities in the mean,
variance, and spatial dependence structure. By isolating stationary components,
spSSA enables their efficient analysis using standard univariate spatial models,
while the nonstationary components—being of reduced dimension—can be modeled
with substantially simplified nonstationary structures.

Unlike much of the existing stationary subspace analysis literature in time
series, where stationary and nonstationary subspace dimensions are usually
assumed to be known, we propose a novel estimator for the latent subspace
dimension. This aspect is of independent interest and not only essential in
practical spatial applications, but also readily transferable to the time series
context.

The remainder of the paper is organized as follows. Section~\ref{sec:ssa}
introduces the necessary definitions and notation as well as presents the spSSA
methods when the nonstationary dimension is known. Section~\ref{sec:rank_est}
presents methods for finding said dimension using data augmentation. These
methods are demonstrated in practice via simulation studies in \Cref{sec:sim1}
and \Cref{sec:sim2}, respectively, as well as using real data in
\Cref{sec:kola}. \Cref{apx:comp,apx:add_plot} show additional simulation results.

\section{Stationary subspace analysis}\label{sec:ssa}

Let $\x(\u)$ be an observable $p$-variate nonstationary random field, $\u \in
\dom \subseteq \R^k$, where $\dom$ denotes the spatial domain. Most commonly the
dimension of the domain is either $k = 2$ or $k = 3$. The methods we present
generalize for arbitrary $k$ but, for our purposes, we fix $k=2$.

Let us further assume that $\x$ can be decomposed into a stationary part and a
nonstationary part as follows:
\begin{equation}
\label{eq::SSAmodel}
  \xu = \A \z + \bs\mu = [\A_{\bo s}\  \A_{\bo n} ] \left(
    \begin{array}{c}
      \s \\
      \n \\
    \end{array}\right) + \bs\mu = \A_{\bo s} \s + \A_{\bo n} \n + \bs\mu,
\end{equation}
where a $p$-variate latent random $\z$ consists of a $(p-q)$-variate stationary
random field $\s$ and a $q$-variate nonstationary random field $\n$, and $\bs
\mu$ is a location vector. Without loss of generality, we can assume that
$\bs\mu = \bs 0$, and will do so henceforth.  
The two components are mixed using a full-rank $p \times p$ matrix $\A$. Here,
matrices $\A_{\bo s}$ and $\A_{\bo n}$ are $p\times (p-q)$ and $p\times q$
matrices, respectively. The aim of spatial stationary subspace analysis (spSSA)
is to estimate an unmixing matrix $\W$ such that $\W\x$ is
partitioned into stationary and nonstationary random fields. 

We make the following assumptions on the latent random fields.
\begin{itemize}
  \item[(A1)] $\E[\s]= \bo 0$, $\Cov(\s) = \bo I$ for all $\bo u$, and
  $\Cov\left(\s,\bo s(\u^\prime)\right)$ is finite and only a function of
  $\|\u-\u^\prime\|= \|\h\|$.
  \item[(A2)] $\E[\n]<\infty$, $\Cov(\n) = \bo D_\u$, where $\bo D_\u$ is a
  diagonal matrix with positive diagonal elements, and $\Cov\left(\n,\bo
  n(\u^\prime)\right)$ is finite.
  \item[(A3)] The random fields $\s$ and $\n$ are independent, i.e., $\Cov(\s, \bo n(\u^\prime)) = \bo 0$ for
  all $\u$ and $\u^\prime$ in $\dom$.
  \item[(A4)] The dimension $q$ is the smallest value that allows division of
  $\x$ such that (A1)-(A3) hold.
\end{itemize}

We thus assume that the $(p-q)$ random fields in $\s$ are second-order
stationary with finite second order spatial dependence. The first assumption
fixes the location and covariance matrix of the stationary part for convenience.
The second assumption states that the $q$ nonstationary components in $\n$ have
finite first and second moments and are uncorrelated at $\u$ but can have
arbitrary location-dependent second order dependence when $\u \neq \u^\prime$.
The third assumption states that nonstationary and stationary components are
independent. The last assumption ensures that $\bo n$ does not include
stationary components.

Despite Assumptions (A1)--(A4), the model is not well-defined. The stationary
components are only specified up to a multiplication by an orthogonal matrix,
whereas the nonstationary components can be marginally rescaled, shifted and
also rotated. Therefore, for convenience of presentation, we make the following
additional assumption on $\n$.
\begin{itemize}
  \item[(A5)] $\sum_{\u \in \U_0} \E[\n]= \bo 0$ and $\sum_{\u \in \U_0} \Cov(\n)
  =  \bo I$, where $\U_0 =\{\u_1,\ldots,\u_{n_0}\} \subset \dom$ is a set of
  $n_0$ reference locations.
\end{itemize}
Hence, based on a set of reference points, location and scale are fixed.
Furthermore, for convenience we will assume that if a realization of $\x$ is
observed at $n$ spatial locations, then these locations will serve as reference
points.

Throughout the rest of this article, we assume that the field
$\x$ is defined as in \Cref{eq::SSAmodel} and satisfies Assumptions
(A1)--(A5). To estimate the unmixing matrix $\W$, we compute
the first and the second order statistics based on the full domain $\dom$ and
then compare these statistics to the ones computed on partitions of the domain.
To this end, let $\dom_1,\ldots,\dom_K$ be a disjoint partition of the domain
$\dom$, that is, $\dom_i \cap \dom_j = \varnothing $ for $1\leq i \neq j \leq K$
and $\bigcup_{k \in K} \dom_k = \dom$. For the stationary signals we should
observe the same statistic locally and globally. Conversely, for the
nonstationary signals the global statistic should differ from the local
statistics.

Suppose that the field $\bo x$ is observed at $n$ locations $\u_1,\ldots, \u_n$.
Define $\U_k=\U_0 \bigcap \dom_k$ to be the set of locations inside the $k$th
subdomain, and let $|\bo U_k|$ denote its cardinality,
for $k=1,\ldots,K$. Denote the local mean, local covariance, and local spatial
covariance, respectively, for subdomain $\U_k$, by
\begin{equation*}
  \begin{split}
    \m_{\U_k}(\x) &= \frac{1}{|\U_k|} \sum_{\u \in\U_k} \xu, \\
    \Cov_{\U_k}(\x) &= \frac{1}{|\U_k|} \sum_{\u\in \U_k}
    \left(\xu - \m_{\U_k}(\x)\right) \left(\xu - \m_{\U_k}(\x)\right)^\top,
    \quad \textnormal{and}\\
    \Lcov_{\U_k,f}(\x) &= \frac{1}{|\U_k|}
    \sum_{\substack{\u,\u^\prime \in\U_k, \\ \u \neq \u^\prime}}
    f(\u-\u^\prime) \left(\xu - \m_{\U_k}(\x)\right)
    \left(\x(\u^\prime) - \m_{\U_k}(\x)\right)^\top,
  \end{split}
\end{equation*}
where $f: \R^2 \rightarrow \R$ denotes a spatial kernel function.
Options, as introduced in \citep{BachocGentonNordhausenRuizGazenVirta2020}, include:

\begin{description}
  \item[Ball kernel:] $f_b(\h;r) = I(\| \h \| \leq r)$, where $r \geq 0$ and $I$
  is the indicator function. \\
  \item[Ring kernel:] $f_r(\h;r_1,r_2) = I(r_1 < \| \h \| \leq r_2)$, where
  $r_1,r_2 \geq 0$ and $r_1 < r_2$.\\
  \item[Gaussian kernel:] $f_g(\h;r) = \exp(-0.5 (\Phi^{-1}(0.95) \| \h \| /
  r)^2)$, where $r > 0$ and $\Phi^{-1}(0.95)$ is the $95\%$ quantile of the
  standard normal distribution.
\end{description}

As is the standard in blind source separation, the first step of the analysis is
to whiten the data \citep{NordhausenRuizGazen2022}. Whitening reduces the
problem to the estimation of an orthogonal transformation, since after
standardization only an orthogonal matrix is required to recover the stationary
and nonstationary subspaces. Accordingly, to separate $\bo s$ and $\bo n$, we
work with the whitened random field defined by
\begin{equation*}
  \xstu = \Cov_{\U}(\x)^{-1/2}(\xu - \bo m_{\U}(\x)).
\end{equation*}

\subsection{Nonstationarity in mean}\label{sec:mean}

Consider first a spSSA method that aims at detecting stationary deviations in
mean. Let $\xst$ be the whitened field corresponding to $\x$ and let 
\begin{equation*}
  \M_\mean = \sum_{k=1}^{K} \frac{|\U_k|}{|\U|}\bo m_{\U_k}(\xst)
  \bo m_{\U_k}(\xst)^\top
\end{equation*}
be the covariance matrix of means computed on different domains weighted by
their number of observations. 

If the mean of a random field does not change in space, then its local means are
approximately the same as the global mean, i.e., zero by
Assumption (A1). The rows and columns, and hence also the diagonals of
$\M_\mean$ corresponding to such fields are zero. Thus, the components that are
stationary in mean, correspond to zero eigenvalues of $\M_\mean$. Conversely,
the components that are nonstationary in mean correspond to non-zero
eigenvalues. Therefore, write the eigenvalue decomposition of $\M_\mean$ as
\begin{equation*}
  \M_\mean = \V_\mean \D_\mean \V_\mean^\top,
\end{equation*}
where $\bo D_\mean$ is a $p\times p$ diagonal matrix of eigenvalues of
$\M_\mean$, and  $\V_\mean$ is the corresponding matrix of eigenvectors. Assume
further that the eigenvalues and vectors are arranged so that $\V_\mean =
(\V_{\mean,s}\ \V_{\mean,n})$ where $\bo V_{\mean,n}$ is the $p\times q_\mean$
matrix containing the eigenvectors of non-zero eigenvalues as columns and the
$p\times (p-q_\mean)$ matrix $\V_{\mean,s}$ includes the remaining columns.
Then, columns of resulting unmixing matrices $\W_{\mean,n} = \V_{\mean,n}^\top
\Cov_{\U}(\x)^{-1/2}$ and $\V_{\mean,s} = \V_{\mean,s}^\top
\Cov_{\U}(\x)^{-1/2}$ generate the nonstationary and stationary subspaces,
respectively. Naturally, $q_\mean \leq q$ with equality only if, for all
nonstationary components, the means differ between at least some of the chosen
partitions. We refer to this method as \textsc{spSSAsir}.

\subsection{Nonstationarity in variance}
\label{sec:var}

Consider again the whitened random field $\xst$. Now 
\begin{equation*}
  \M_\var = \sum_{k=1}^{K} \frac{|\U_k|}{|\U|}
  \left(\bo I- \Cov_{\U_k}(\xst)\right)^2,
\end{equation*}
where $\bo A^2=\bo A\bo A^\top$, measures the deviation of the covariance matrix
computed on partitions $\U_1,\ldots,\U_K$ from the global covariance matrix $\bo
I_p$. 

Again, the eigenvalues of $\M_\var$ that correspond to the components that are
nonstationary in variance should be non-zero. Let the eigenvalue decomposition
be
\begin{equation*}
  \M_\var = \V_\var \D_\var \V_\var^\top,
\end{equation*}
where $\D_\var$ is a $p\times p$ diagonal matrix of eigenvalues of $\M_\var$,
and $\V_\var = (\V_{\var,s}\ \V_{\var,n})$ are the eigenvectors of $\M_\var$
arranged so that $\V_{\var,n}$ is the $p\times q_\var$ matrix containing the
eigenvectors corresponding to the non-zero eigenvalues as columns and
$\V_{\var,s}$ is the $p\times (p-q_\var)$ matrix containing the remaining
eigenvectors as columns. Again $q_\var \leq q$. The transformations to the two
subspaces are accordingly given by the matrices $\W_{\var,n} = \V_{\var,n}^\top
\Cov_{\U}(\xst)^{-1/2}$ and $\W_{\var,s} = \V_{\var,s}^\top
\Cov_{\U}(\xst)^{-1/2}$. Although this approach is designed to detect components
with nonstationary variances, it may also detect components which are
nonstationary in mean as, if the mean changes, the variances in partitions might
also change. We refer to this method as \textsc{spSSAsave}.

\subsection{Nonstationarity in spatial dependence}
\label{sec:cor}

The two previous methods are able to detect components which are nonstationary
with respect to the first and the second moments. To detect nonstationarities in
spatial dependence, we need a statistic that measures the
variability of spatial dependence across subdomains. 
Let such a statistic be defined (for whitened fields) as  
\begin{equation*}
  \M_{\cor,f} = \sum_{k=1}^{K} \frac{|\U_k|}{|\U|}
  \left(\Lcov_{\U,f}(\xst)- \Lcov_{\U_k,f}(\xst)\right)^2.
\end{equation*}
 The matrix $\M_{\cor,f}$ measures the deviation of the local covariance
 matrices computed on partitions $\U_1,\dots, \U_K$ from the global local
 covariance matrix $\Lcov_{\U,f}$. Using again the eigendecomposition
\begin{equation}
  \label{eq:Vcor}
  \M_{\cor,f} = \V_\cor \bo D_\cor \V_\cor^\top,
\end{equation}
and separating the eigenvectors of $\M_{\cor,f}$ corresponding to non-zero and
zero eigenvalues yields the $p\times q_{\cor}$ matrix $\V_{\cor,n}$ and the
$p\times (p-q_{\cor})$ matrix $\V_{\cor,s}$, where $q_\cor$ is the number of
non-zero eigenvalues with  $q_\cor \leq q$. The unmixing matrix estimates are
$\W_{\cor,n} = \V_{\cor,n}^\top \Cov_{\U}(\xst)^{-1/2}$ and $\W_{\cor,s} =
\V_{\cor,s}^\top \Cov_{\U}(\xst)^{-1/2}$. The method is able to
detect nonstationary components if at least some subdomains
exhibit differences in their local covariance structure. As this method is for
detecting changes in the spatial correlation structure we denote it as
\textsc{spSSAcor}. 

Notice that as $\M_{\cor,f}$ is computed using a single kernel $f$, the
performance of \textsc{spSSAcor} will naturally depend on the choice of $f$.
However, there is no obvious rule for selecting $f$ in practice in the absence
of additional information about the underlying processes. Some visual and
exploratory guidelines for choosing $f$ in the context of SBSS are discussed in
\citet{PiccolottoEtal2022}. An alternative approach is to employ multiple
kernels $f_1,\ldots,f_L$ and to perform joint diagonalization of
$\M_{\cor,f_1},\dots,\M_{\cor,f_L}$, rather than solely relying on the
eigendecomposition in \Cref{eq:Vcor}.

Note that all the three preceding methods are formulated as an eigenvalue
problem of a matrix computed from the whitened data $\xst$. Equivalently, they
can be formulated as generalized eigenvalue problems for the matrix pair
$(\M\Cov_{\U}(\x)^{-1/2}, \Cov_{\U}(\x)^{1/2})$.

\subsection{Combination of methods}
\label{sec:comb}

Denote from now on $\M_\mean= \M_1$, $\M_\var= \M_2$ and $\M_{\cor,f}= \M_3$.
The methods of Sections \Cref{sec:mean,sec:var,sec:cor} can be combined to detect all three types of nonstationarities by solving a joint diagonalization problem. 
Since exact joint diagonalization of
more than two matrices is usually impossible, approximate joint diagonalization
is required. For the three matrices $\bo M_i$, $i=1,2,3$, computed with the
whitened random field, this amounts to finding an orthogonal $p\times p$ matrix
$\bo V_\comb$ that minimizes  
$\sum_{i=1}^3 \left\|\off(\V_\comb^\top \M_i \bo V_\comb )\right\|^2$, or,
equivalently, since $\V_\comb$ is orthogonal, maximizes
\begin{equation}
  \label{eq:jointdiag}
  \sum_{i=1}^3 \left\|\diag(\V_\comb^\top \M_i \V_\comb )\right\|^2.
\end{equation}
Here $\left\|\bo A\right\|$ is the matrix (Frobenius) norm, $\diag(\bo A)$ is a
$p\times p$ diagonal matrix with the diagonal elements as in $\bo A$ and
$\off(\bo A) = \bo A - \diag(\bo A)$. Notice that the principle of the
estimation procedure remains the same even if additional matrices
$\M_{\cor,f_i}$ with various kernels $f_i$ are included in the objective
function. Similarly, it is possible to only consider a subset of the matrices
$\M_i$.

Several algorithms for approximate joint diagonalization in~\Cref{eq:jointdiag}
exist in the literature. The most popular one based on Givens rotations is
proposed in \cite{Clarkson:1988} and is available for example in the R package
JADE \citep{JADE_package}. 

Based on $\V_\comb$, one can compute $\D_i = \diag(\V_\comb \M_i \V^\top_\comb
)=\diag(d_{i,1},\ldots, d_{i,p})$ and collect all those columns $j$ of
$\V_\comb$, where $\sum_i |d_{i,j}| \neq 0$, to a $p\times q_\comb$ matrix
$\V_{\comb,n}$. The rest of the columns are then collected to a $p\times
(p-q_\comb)$ matrix $\V_{\comb,s}$. Here the individual value $d_{i,j}$
indicates whether the $j$th component is nonstationary with respect to $\bo
M_i$. The final transformation matrices for nonstationary and stationary
components are then $\W_{\comb,n} = \V^\top_{\comb,n} \Cov_{\U}(\xst)^{-1/2}$
and $\W_{\comb,s} = \V^\top_{\comb,s} \Cov_{\U}(\xst)^{-1/2}$. We refer to this
method as \textsc{spSSAcomb}.

\begin{remark}\label{remark:scale} Alternatively, one may consider forming a
linear combination of the matrices $\M_1, \M_2,$ and $\M_3$ and performing an
eigenvalue decomposition of the resulting matrix. This approach, however,
introduces nontrivial scaling issues. For example, in a time series context,
\citet{Haraetal:2010} consider a weighted linear combination of $\M_1$ and
$\M_2$. Such weighting becomes necessary when one of the matrices has a
substantially larger norm than the others; without appropriate scaling, the
combined matrix would yield results that differ only marginally from those
obtained using the dominant matrix alone. A seemingly straightforward remedy is
to scale each matrix by an appropriate matrix norm. However, this can be
problematic: if a particular type of nonstationarity is absent, the
corresponding scatter matrix contains only noise, and dividing by its nonzero,
but very small norm, may strongly amplify the noise. For this reason, we do not
consider this approach in the simulation study. We note that similar scaling
issues may also arise in the context of approximate joint diagonalization.
\end{remark}

\begin{remark}\label{remark:SDR} As noted by
 \citet{FlumianMatilainenNordhausenTaskinen2024} in the time series setting, the
 SSA approaches considered here may also be viewed through the lens of
 supervised dimension reduction (SDR). From this perspective, an artificial
 response variable is given by the partition membership, and the methods aim to
 identify components that best discriminate between partitions, that is, the
 nonstationary components. This interpretation also explains the terminology SIR
 and SAVE, which is motivated by the corresponding SDR methods sliced inverse
 regression \citep{Li:1991} and sliced average variance estimation
 \citep{Cook:2000}.
\end{remark}

\subsection{Practical considerations}

\begin{figure}[ht!]
    \begin{subfigure}[b]{0.49\textwidth}
        \centering
        \includegraphics[width=\linewidth]{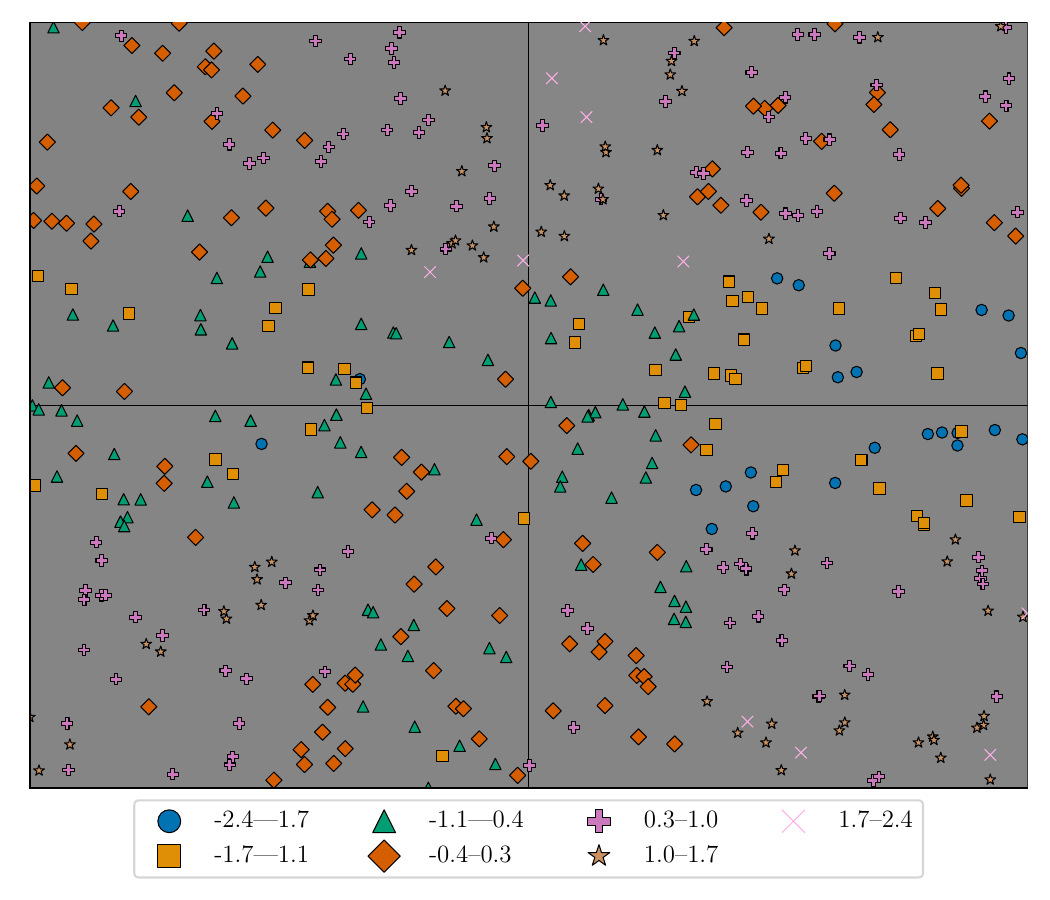}
        \caption{2-by-2 partition}
        \label{fig:part_subfig2}
    \end{subfigure}
    \begin{subfigure}[b]{0.49\textwidth}
        \centering
        \includegraphics[width=\linewidth]{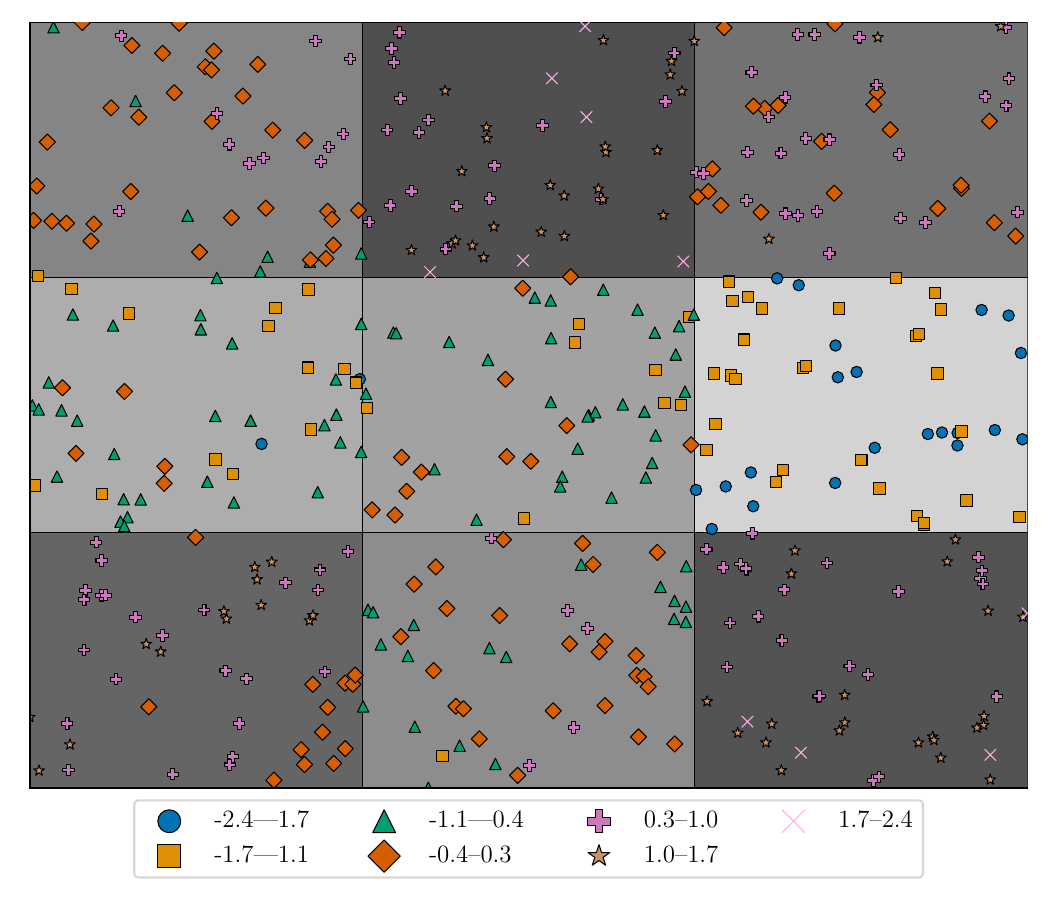}
        \caption{3-by-3 partition}
        \label{fig:part_subfig3}
    \end{subfigure}
    \caption{Example of the importance of choosing the right partition for a
    spatial signal that is nonstationary in mean. The labels represent ranges of
    observed values of the spatial signal. The background color represents the
    mean of a box. The darker the color the higher the mean.}
    \label{fig:partition_example}
\end{figure}

The choice of the spatial partition plays a crucial role in the performance of
the spSSA procedures. An unsuitable partition may cause the methods to indicate
the absence of nonstationarity, even when pronounced local differences are
present. This is closely related to the \textit{modifiable areal unit problem}
(MAUP), see e.g., \cite{Openshaw1984, FotheringhamWong1991}.  An illustrative
example is shown in \Cref{fig:partition_example}, which displays two subfigures
with the same univariate random field with 450 data points. The range of values
is split into seven groups, and the value of the random field at a given
location is represented by the label of the group. The ranges and labels are given
in the legend. Globally, the data has mean 0 and variance 1. While the random field is
stationary in variance and covariance, its mean varies locally across the
domain. The spatially varying mean is generated on a $3 \times 3$ grid, and
using a corresponding $3 \times 3$ partition allows \textsc{spSSAsir} to
successfully detect this nonstationary structure. In contrast, when a $2 \times
2$ partition is used, the regional means do not differ, and the nonstationarity
in the mean remains undetected. 

\begin{figure}[ht!]
    \begin{subfigure}[b]{0.49\textwidth}
        \centering
        \includegraphics[width=\linewidth]{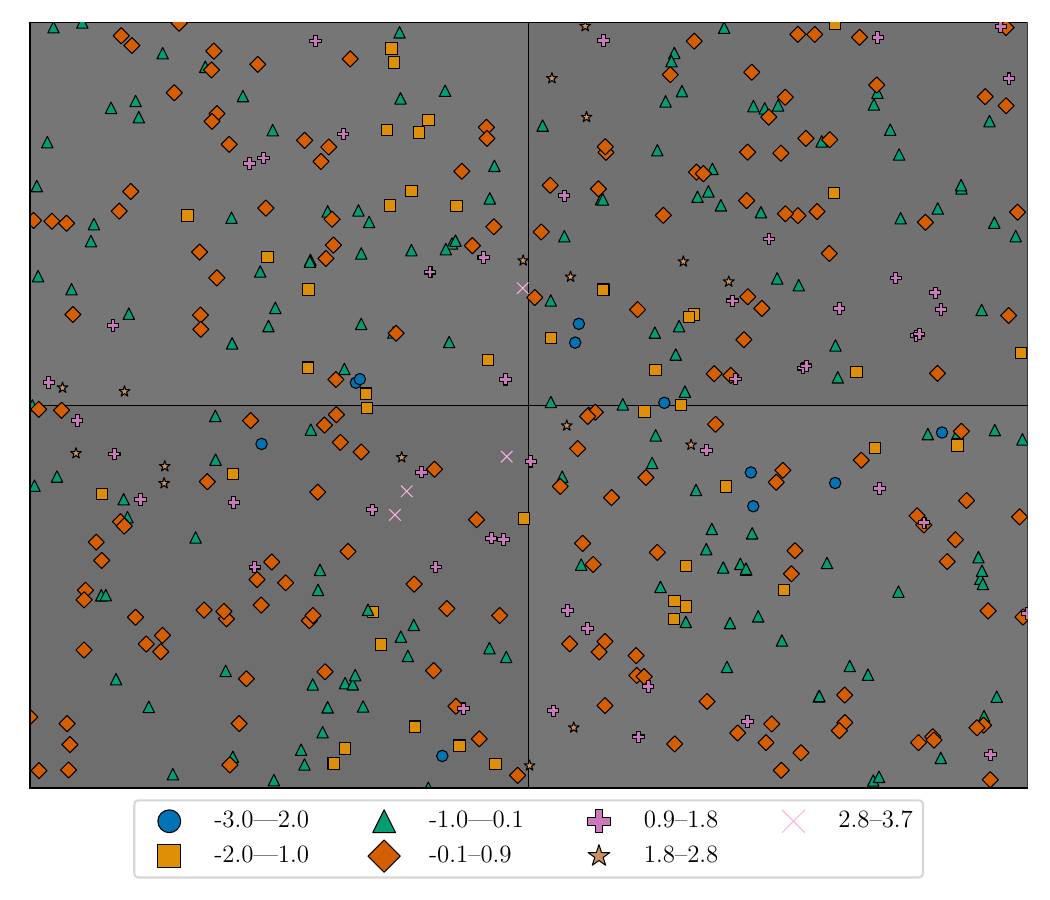}
        \caption{2-by-2 partition}
        \label{fig:part_var_subfig2}
    \end{subfigure}
    \begin{subfigure}[b]{0.49\textwidth}
        \centering
        \includegraphics[width=\linewidth]{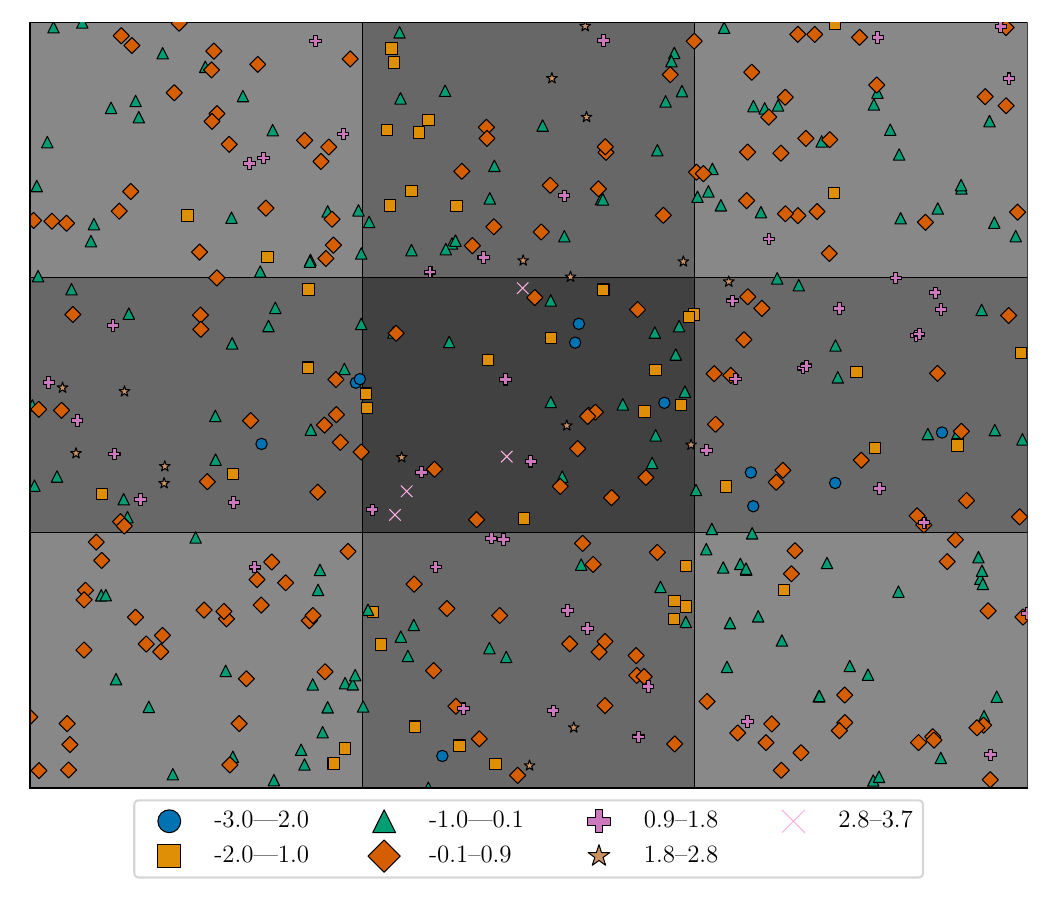}
        \caption{3-by-3 partition}
        \label{fig:part_var_subfig3}
    \end{subfigure}
    \caption{Example of the importance of choosing the right partition for a
    spatial signal that is nonstationary in variance. The labels represent
    ranges of observed values of the spatial signal. The background color
    represents the variance of a box. The darker the color the higher the
    variance.}
    \label{fig:partition_var_example}
\end{figure}

A similar effect is illustrated for variance in
\Cref{fig:partition_var_example}, where this time the random field is stationary
in mean but nonstationary in variance.  
Analogous issues arise for nonstationarity in the spatial dependence structure.
In general, partitions should be chosen to maximize differences between regions,
a task for which subject-matter knowledge can be highly beneficial.

As noted in \Cref{remark:scale}, the relative scaling of the $\M$ matrices can
substantially affect the performance of \textsc{spSSAcomb}. To fully exploit
approximate joint diagonalization, it is important that the norms of these
matrices are of comparable magnitude. If one matrix dominates in norm, the
resulting decomposition would essentially recover the eigenstructure of the
dominant matrix, with only a negligible contribution from the remaining
matrices. 

After whitening the signals, the matrices $\M_{\mean}$ and $\M_{\var}$ are
expected to be of comparable norm. This is not necessarily the case for
$\M_{\cor}$. For purely stationary signals, local means and variances should
closely match their global counterparts, implying that the norms of $\M_{\mean}$
and $\M_{\var}$ remain small. In contrast, local and global spatial covariances
can differ significantly, even under stationarity, due to variations in the
spatial configuration and density of measurement locations within partitions.
This effect becomes more pronounced depending on the partition, and the norm of
$\M_{\cor,f}$ can potentially grow alongside the number of partitions. 

This motivates the introduction of scaling. By appropriately dampening the
influence of densely sampled regions, scaling reduces exaggerated local--global
differences in spatial dependence and keeps the norm of $\M_{\cor,f}$ on a scale
comparable to that of $\M_{\mean}$ and $\M_{\var}$.

Let us define a scaled version of local covariance $\Lcov^s_{\U_k,f}$ by
\begin{equation*}
    \Lcov^s_{\U_k,f}(\x) = \frac{1}{|\U_k|}
    \sum_{\u\in\U_k} \frac{1}{F_{\U_k}(\u;f)} \sum_{\u' \in \U_k\setminus \u} f(\u-\u^\prime) \left(\xu - \m_{\U_k}(\x)\right)
    \left(\x(\u^\prime) - \m_{\U_k}(\x)\right)^\top,
\end{equation*} where \[
    F_{\U_k}(\u; f) = \sum_{\u' \in \U_k\setminus \u} f(\u - \u').
\] Analogously, define $\M^s_{\cor, f}$ by \[
    \M^s_{\cor,f} = \sum_{k=1}^{K} \frac{|\U_k|}{|\U|}
  \left(\Lcov^s_{\U,f}(\xst)- \Lcov^s_{\U_k,f}(\xst)\right)^2.
\] 

The scaling factor accounts for the number of nearby locations contributing to
the local covariance at a location $\u$. For the ball kernel, the scaling factor
$F_{\U_k}(\u,f_b(\,\cdot\,;r))$ is simply the number of measurement locations
contained in $B(\u,r)\cap\U_k$. The scaled local covariance retains the same
structural properties as the unscaled version, since the scaling factor can be
absorbed into the kernel function.

The use of a scaled local covariance is not strictly required for any of the
methodologies introduced in this paper. However, in the context of approximate
joint diagonalization, it resolves the scaling and partitioning issues discussed
above. In particular, the scaled local covariance allows $\M_{\comb}$ to capture
local covariance information without overwhelming the contributions of
$\M_{\mean}$ and $\M_{\var}$. For this reason, without loss of generality, we
employ the scaled local covariance estimator in the subsequent sections, while
continuing to refer to the corresponding method as \textsc{spSSAcor}. Note that
there are also other scaled variants of the local covariance, as for example
discussed in \citet{MuehlmannBachocNordhausenYi2024}. A comparison between the
scaled and unscaled \textsc{spSSAcor}, as well as an example of the scaling
issue, can bee seen in \Cref{apx:comp}.

\section{Simulations I}\label{sec:sim1}

In this section, we assess the performance of the proposed methods under the
assumption that the subspace dimension $q$ is known. For a given unmixing matrix
$\W = \left[\W_s \; \W_n \right]$ and its estimate $\widetilde{\W} =
\left[\widetilde{\W}_s \; \widetilde{\W}_n \right]$, we compute their respective
projectors 
\[
    \bo P_s = \W_s \left(\W_s^\top \W_s \right)^{-1} \W_s^\top,
    \hspace{1cm}\bo P_n = \W_n \left(\W_n^\top \W_n \right)^{-1} \W_n^\top
\]
and 
\[
    \widetilde{\bo P}_s = \widetilde{\W}_s\left(\widetilde{\W}_s^\top
    \widetilde{\W}_s \right)^{-1} \widetilde{\W}_s^\top,\hspace{1cm}
    \widetilde{\bo P}_n = \widetilde{\W}_n\left(\widetilde{\W}_n^\top
    \widetilde{\W}_n \right)^{-1} \widetilde{\W}_n^\top.
\]
We choose, as a performance metric, the squared distance between the true and
estimated projectors. That is, we denote by $\bo s_{\text{perf}}$ the stationary
performance and $\bo n_{\text{perf}}$ the nonstationary performance, which are
given by 
\[
\bo s_{\text{perf}} = \frac{1}{2} \norm{\bo P_s - \widetilde{\bo P}_s}^2 \;
\text{ and }\;
\bo n_{\text{perf}} = \frac{1}{2} \norm{\bo P_n - \widetilde{\bo P}_n}^2
\]
For properties of the performance metric, see
\cite{LiskiNordhausenOjaRuizGazen2016}. A key result is that the performance
metric is upper bounded by the dimension of the subspace we are projecting to,
that is, $\bo s_{\text{perf}} \in [0, \dim(\s)]$ and ${\bo n_\text{perf}} \in
[0, \dim(\n)]$. Ideally, we want these metrics to converge to zero as we
increase the number of measurements.

All of the following simulations are done with Python, and the source code is
available at \url{github.com/perttusaarela/sp_ssa/releases/tag/v1.0}. In all
settings, the observed random field $\xu$ is 8-dimensional with the stationary
subspace $\s$ begin 5-dimensional and the nonstationary subspace $\n$ having
dimension 3. The observed locations are drawn uniformly at random from the
domain $\dom = [0, \ell]^2$, that is, a square area whose side length is $\ell
\in \{20, 30, 40, 50, 60, 70\}$. The number of locations $n$ is the side length
squared. In this way, we keep approximately the same density as we increase the
number of measurements. The latent components are mixed by a random $8\times 8$
orthogonal matrix $\A$. This is done without loss of generality, since the data
is standardized. 

For the spatial dependencies between these measurement locations, we use the
Mat\'ern covariance matrix defined by the function
\[
        C(\bo h; \nu, \phi) = \frac{1}{2^{\nu-1}\Gamma(\norm{\bo h})}
        \left(\frac{\norm{\bo h}}{\phi}\right)^{\nu}K_\nu
        \left(\frac{\norm{\bo h}}{\phi}\right),
\]
where $\norm{\,\cdot\,}$ is the Euclidean norm, $K_\nu$ is a modified Bessel
function of the second-kind, $\nu$ is the shape parameter and $\phi$ is the
range parameter. The vector $\bo h$ is the difference between two measurement
points. 

Components $\bo s_i(\bo u),$ $i \in \{1, 2, 3, 4, 5\}$, of the
stationary signal $\s$ are independent zero mean Gaussian processes, where the
covariance structure is determined by the Mat\'ern covariance function
\[
\Cov(\bo s_i (\bo u), \bo s_i(\bo u')) = C(\bo u - \bo u';
0.5, 1.0),
\]
which guarantees that Assumption (A1) is satisfied. For the
signals $\n$, we consider three types of nonstationarities. In the first three
settings, only one type of nonstationarity is present. Settings 1, 2, and 3 have
nonstationarity in mean, variance, and spatial covariance, respectively. Setting
4 will combine all of these nonstationarities.

All the methods above rely on a partition $\dom_1,\ldots,\dom_K$ of the domain.
For the simulations, we use a simple grid partition. For each of the settings,
we consider three partitions: 2-by-2, 3-by-3, and 4-by-4 (so $K=4, 9, 16$
respectively). In subsequent plots a $k$-by-$\ell$ partition is denoted by $(k,
\ell)$. We compute the averages of performance measures from
2000 trials to smooth out single run variability. 

Finally, along with the introduced methods, we also plot the performance of a
random baseline. This is computed by generating a random $n\times n$ orthogonal
matrix $\U$ and setting the unmixing matrix to $ \W =
\U^\top\Cov_{\U}(\xst)^{-1/2}$. The idea of the baseline is to demonstrate the
efficacy of our methods compared to random guessing.

\paragraph{Setting 1} 
The stationary signal $\s$ is defined as above. The nonstationary signal $\bo n
(\bo u)$ consists of three components of the form $\bo n_i(\bo u) = y_i(\bo u) +
\mu_i(\bo u)$, where $y_i(\bo u)$ is a stationary signal generated in the same way
as the components of $\s$, and $\mu_i(\bo u)$ is a spatially varying mean
function. The function $\mu_i$ is a piecewise constant function whose values are
defined by the partition $A^{(i)}$. The mean functions are 

\begin{minipage}{0.32\textwidth}
    \begin{equation*}
    \mu_1(\bo u) = \begin{cases}
        1.5 &\bo u \in A^{(1)}_1 \text{ or } A^{(1)}_4 ,\\
        -1.5 & \bo u \in A^{(1)}_2 \text{ or } A^{(1)}_3,\\
    \end{cases}
\end{equation*}
\end{minipage}
\begin{minipage}{0.32\textwidth}
 \begin{equation*}
    \mu_2(\bo u) = \begin{cases}
        1 &\bo u \in A^{(2)}_1 \text{ or } A^{(1)}_5,\\
        -0.5 & \bo u \in A^{(2)}_2 \text{ or } A^{(1)}_6,\\
        2.0 & \bo u \in A^{(2)}_3 \text{ or } A^{(1)}_4,\\
    \end{cases}
\end{equation*}
\end{minipage}
\begin{minipage}{0.32\textwidth}
    \begin{equation*}
    \mu_3(\bo u) = \begin{cases}
        -1.5 &\bo u \in A^{(3)}_1 \text{ or } A^{(1)}_6,\\
        -0.5 & \bo u \in A^{(3)}_2 \text{ or } A^{(1)}_7,\\
        0.5 & \bo u \in A^{(3)}_3 \text{ or } A^{(1)}_8,\\
        1.5 & \bo u \in A^{(3)}_4 \text{ or } A^{(1)}_5.
    \end{cases}
\end{equation*}
\end{minipage}

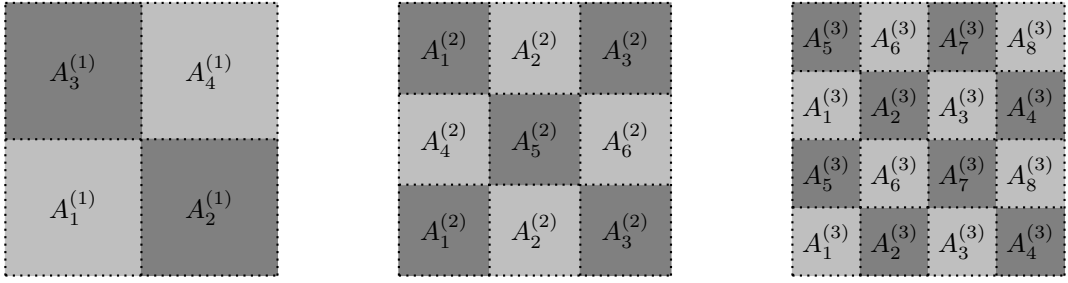
\begin{figure}[ht!]
    \begin{subfigure}[b]{0.31\textwidth}
    \centering
    \begin{tikzpicture}[scale=1.2]
            \path[use as bounding box] (0,0) rectangle (3,3);
    \draw [draw=lightgray, fill=lightgray]  (0,0) rectangle (1.5,1.5) node[midway] {$A^{(1)}_1$};
    \draw [draw=lightgray, fill=lightgray]  (1.5,1.5) rectangle (3.0,3.0) node[midway] {$A^{(1)}_4$};
    \draw [draw=gray, fill=gray]  (1.5,0) rectangle (3,1.5) node[midway] {$A^{(1)}_2$};
    \draw [draw=gray, fill=gray]  (0,1.5) rectangle (1.5,3) node[midway] {$A^{(1)}_3$};
    \draw[black, dotted, thick, step=1.5] (0, 0) grid (3, 3);
\end{tikzpicture}
        \caption{Partition of parameters for signals $n_1$}
        \label{fig:subfig8}
    \end{subfigure}
    \begin{subfigure}[b]{0.31\textwidth}
    \centering
    \begin{tikzpicture}[scale=1.2]
            \path[use as bounding box] (0,0) rectangle (3,3);
    \draw [draw=gray, fill=gray]  (0,0) rectangle (1,1) node[midway] {$A^{(2)}_1$};
    \draw [draw=lightgray, fill=lightgray]  (1,0) rectangle (2,1) node[midway] {$A^{(2)}_2$};
    \draw [draw=gray, fill=gray]      (2,0) rectangle (3,1) node[midway] {$A^{(2)}_3$};
    \draw [draw=lightgray, fill=lightgray]  (0,1) rectangle (1,2) node[midway] {$A^{(2)}_4$}; 
    \draw [draw=gray, fill=gray]  (1,1) rectangle (2,2) node[midway] {$A^{(2)}_5$};
    \draw [draw=lightgray, fill=lightgray]      (2,1) rectangle (3,2) node[midway] {$A^{(2)}_6$};
    \draw [draw=gray, fill=gray]  (0,2) rectangle (1,3) node[midway] {$A^{(2)}_1$}; 
    \draw [draw=lightgray, fill=lightgray]  (1,2) rectangle (2,3) node[midway] {$A^{(2)}_2$};
    \draw [draw=gray, fill=gray]      (2,2) rectangle (3,3) node[midway] {$A^{(2)}_3$};
    \draw[black, dotted, thick] (0, 0) grid (3, 3);
\end{tikzpicture}
        \caption{Partition of parameters for signals $n_2$}   
        \label{fig:subfig9}
    \end{subfigure}
    \begin{subfigure}[b]{0.31\textwidth}
    \centering
    \begin{tikzpicture}[scale=1.2]
            \path[use as bounding box] (0,0) rectangle (3,3);
    \draw [draw=lightgray, fill=lightgray]  (0,0) rectangle (0.75,0.75) node[midway] {$A^{(3)}_1$};
    \draw [draw=gray, fill=gray]  (0.75,0) rectangle (1.5,0.75) node[midway] {$A^{(3)}_2$};
    \draw [draw=lightgray, fill=lightgray]     (1.5,0) rectangle (2.25, 0.75) node[midway] {$A^{(3)}_3$};
    \draw [draw=gray, fill=gray]   (2.25,0) rectangle (3,0.75) node[midway] {$A^{(3)}_4$};
    \draw [draw=gray, fill=gray]   (0,0.75) rectangle (0.75,1.5) node[midway] {$A^{(3)}_5$}; 
    \draw [draw=lightgray, fill=lightgray]   (0.75,0.75) rectangle (1.5,1.5) node[midway] {$A^{(3)}_6$}; 
    \draw [draw=gray, fill=gray]      (1.5,0.75) rectangle (2.25, 1.5) node[midway] {$A^{(3)}_7$};
    \draw [draw=lightgray, fill=lightgray]    (2.25,0.75) rectangle (3,1.5) node[midway] {$A^{(3)}_8$};
    \draw [draw=lightgray, fill=lightgray]  (0,1.5) rectangle (0.75,2.25) node[midway] {$A^{(3)}_1$}; 
    \draw [draw=gray, fill=gray]  (0.75,1.5) rectangle (1.5,2.25) node[midway] {$A^{(3)}_2$}; 
    \draw [draw=lightgray, fill=lightgray]     (1.5,1.5) rectangle (2.25, 2.25) node[midway] {$A^{(3)}_3$};
    \draw [draw=gray, fill=gray]   (2.25,1.5) rectangle (3,2.25) node[midway] {$A^{(3)}_4$};
    \draw [draw=gray, fill=gray]   (0,2.25) rectangle (0.75,3) node[midway] {$A^{(3)}_5$}; 
    \draw [draw=lightgray, fill=lightgray]   (0.75,2.25) rectangle (1.5,3) node[midway] {$A^{(3)}_6$};
    \draw [draw=gray, fill=gray]     (1.5,2.25) rectangle (2.25, 3) node[midway] {$A^{(3)}_7$};
    \draw [draw=lightgray, fill=lightgray]    (2.25,2.25) rectangle (3,3) node[midway] {$A^{(3)}_8$};
    \draw[black, dotted, thick, step=0.75cm] (0, 0) grid (3, 3);
\end{tikzpicture}
        \caption{Partition of parameters for signals $n_3$}
        \label{fig:subfig10}
    \end{subfigure}
    \caption{Partitions for parameters used for generating the nonstationary signal $\n$ in all settings.}
    \label{fig:partitions}
\end{figure}

\Cref{fig:partitions} shows the partitions $A^{(i)}$ which are reused for all settings. 
More concretely, the defining partition of a nonstationary signal $\bo n_i$ is a $(i+1)$-by-$(i+1)$ grid partition.
To accentuate the results, we assign
different parameters to neighboring parts of the partition. To this end, we
reuse parameters for multiple parts. Note that these are the same partitions
that we also use for the estimations.

\begin{figure}
    \centering
    \includegraphics[width=0.8\linewidth]{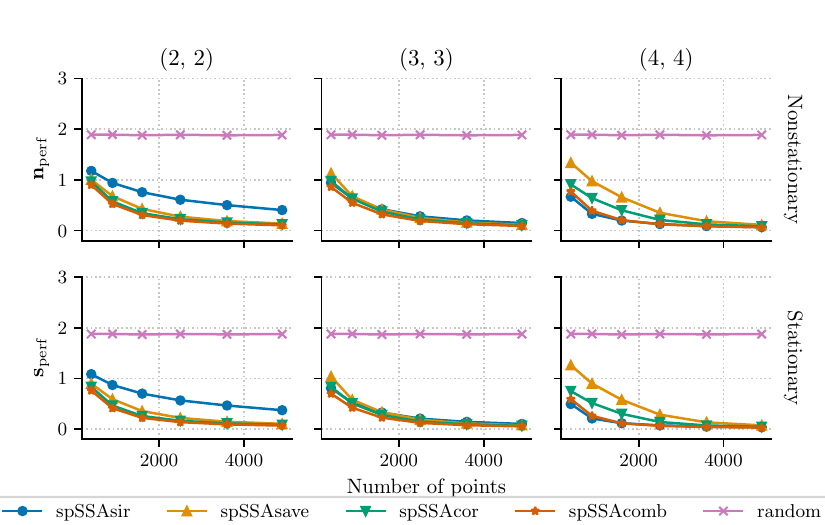}
    \caption{Setting 1: nonstationarity in the mean.}
    \label{fig:sim1_setting1}
\end{figure}

\Cref{fig:sim1_setting1} shows that all the presented methods find
nonstationarities in the mean. Based on how the signals are defined,
\textsc{spSSAsir} struggles slightly on the 2-by-2 partition but outperforms
other methods when granularity is increased.

\paragraph{Setting 2}
The stationary part $\s$ is as defined above. The nonstationary part $\bo n
(\u)$ consists of three signals $\bo n_i$, where each signal is generated by
scaling a stationary signal $y$ with a regionally varying standard deviation. In
other words, $\bo n_i(\u) = \sigma_i (\u) y_i(\u)$, where $y_i$ is generated as in
Setting 1 and the scaling functions are defined as 
\begin{minipage}{0.32\textwidth}
    \begin{equation*}
    \sigma_1^2(\bo u) = \begin{cases}
        0.4 &\bo u \in A^{(1)}_1 \text{ or } A^{(1)}_4 ,\\
        1.4 & \bo u \in A^{(1)}_2 \text{ or } A^{(1)}_3,\\
    \end{cases}
\end{equation*}
\end{minipage}
\begin{minipage}{0.32\textwidth}
\begin{equation*}
    \sigma_2^2(\bo u) = \begin{cases}
        3.0 &\bo u \in A^{(2)}_1 \text{ or } A^{(1)}_5,\\
        0.5 & \bo u \in A^{(2)}_2 \text{ or } A^{(1)}_6,\\
        1.5 & \bo u \in A^{(2)}_3 \text{ or } A^{(1)}_4,\\
    \end{cases}
\end{equation*}
\end{minipage}
\begin{minipage}{0.32\textwidth}
\begin{equation*}
    \sigma_3^2(\bo u) = \begin{cases}
        0.4 &\bo u \in A^{(3)}_1 \text{ or } A^{(1)}_6,\\
        0.8 & \bo u \in A^{(3)}_2 \text{ or } A^{(1)}_7,\\
        1.5 & \bo u \in A^{(3)}_3 \text{ or } A^{(1)}_8,\\
        1.2 & \bo u \in A^{(3)}_4 \text{ or } A^{(1)}_5.
    \end{cases}
\end{equation*}
\end{minipage}

The regions for each signal are the same as the corresponding signal in Setting
1 and can be seen in \Cref{fig:partitions}. This method for generating the data
is identical to using the following  covariance matrix for $\bo n_k(\u)$: 
\[
    \Cov(\bo n_k(\bo u_i), \bo n_k(\bo u_j)) =\sigma_k(\bo u_i)\sigma_k(\bo u_j)
    C(\bo u_i - \bo u_j; 0.5, 1.0).
\]
Note that this also causes some nonstationarity in the spatial covariance.

\begin{figure}
    \centering
    \includegraphics[width=0.8\linewidth]{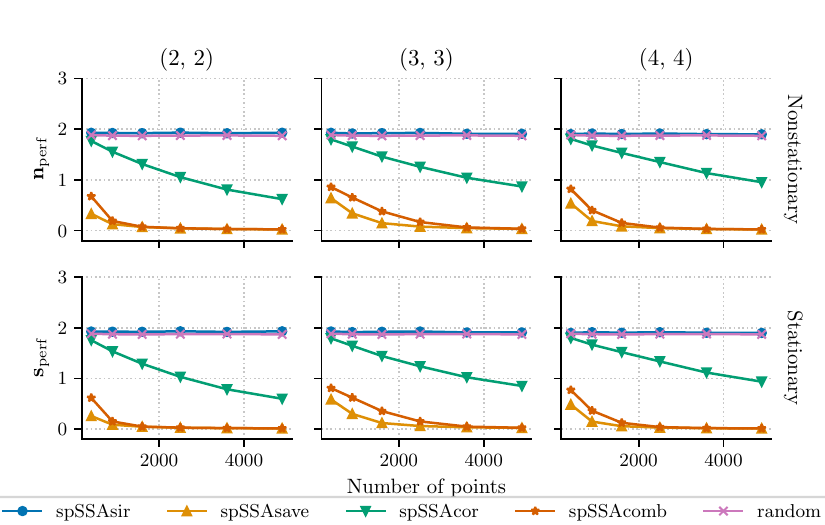}
    \caption{Setting 2: nonstationarity in variance}
    \label{fig:sim1_setting2}
\end{figure}

\Cref{fig:sim1_setting2} shows that \textsc{spSSAsave} finds all the signals
even with few measurements. Since the mean is zero both globally and locally,
\textsc{spSSAsir} performs as bad as the baseline. With enough data points,
\textsc{spSSAcor} seems to be able to find some nonstationary signals. This is
to be expected based on how we actually generate the data.

\paragraph{Setting 3}
The stationary signals are again generated as above. For the nonstationary
signals, each signal is generated independently. For each signal $\bo n_i(\u)$,
data are generated independently across partitions using Matérn covariances with
partition-specific parameters $(\nu, \phi)$ (see \Cref{tab:simI}). 

\begin{table}[ht]
\centering
\caption{Parameter values $(\nu,\phi)$ for each block in the three partitions.}
\label{tab:simI}
\begin{tabular}{lcccccccc}
\toprule
Partition 
& $A^{(\ell)}_1$ 
& $A^{(\ell)}_2$ 
& $A^{(\ell)}_3$ 
& $A^{(\ell)}_4$ 
& $A^{(\ell)}_5$ 
& $A^{(\ell)}_6$ 
& $A^{(\ell)}_7$ 
& $A^{(\ell)}_8$ \\
\midrule
$\ell=1$ 
& (0.3, 0.5) 
& (1.5, 1.3) 
& (1.0, 2.0) 
& (0.5, 2.0) 
& -- 
& -- 
& -- 
& -- \\

$\ell=2$ 
& (1.0, 1.5) 
& (0.5, 0.8) 
& (2.0, 1.7) 
& (0.5, 2.0) 
& (1.0, 2.0) 
& (0.5, 2.0) 
& -- 
& -- \\

$\ell=3$ 
& (1.6, 1.6) 
& (0.3, 0.3) 
& (2.5, 3.0) 
& (0.8, 3.0) 
& (0.5, 1.8) 
& (1.0, 3.0) 
& (0.5, 1.2) 
& (0.3, 2.5) \\
\bottomrule
\end{tabular}
\end{table}

We use the ball kernel $f_b(\bo h; r)$ with radius $r = 3.4$ for all
simulations. The radius is chosen such that we have sufficient number of
observations to estimate the covariance matrices. Assuming uniform density for
the points, there should be approximately $\pi r^2$ points in a ball with radius
$r$. Since we are estimating an $8\times 8$ covariance matrix, we have 36
parameters to estimate. Setting $r = 3.4$ gives $\pi r ^2 \approx 36.3$.

\begin{figure}
    \centering
    \includegraphics[width=0.8\linewidth]{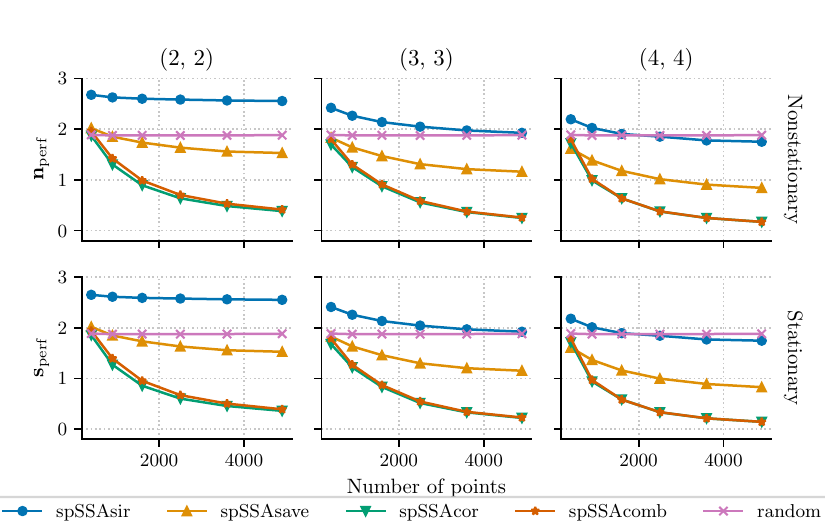}
    \caption{Setting 3: nonstationarity in spatial covariance.}
    \label{fig:sim1_setting3}
\end{figure}

\Cref{fig:sim1_setting3} shows that \textsc{spSSAsir} finds nothing,
while \textsc{spSSAsave} finds some signals with enough data points. The best
method is, as expected, \textsc{spSSAcor} followed by \textsc{spSSAcomb}. This
also highlights the importance of the partition: even \textsc{spSSAcor} does not
fully converge to zero when the partition is too sparse. 

\begin{figure}
    \centering
    \includegraphics[width=0.8\linewidth]{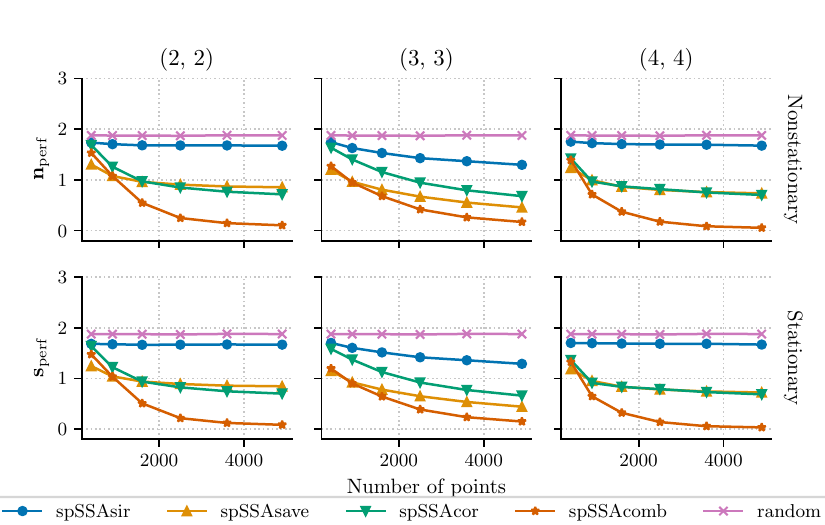}
    \caption{Setting 4: all three types of nonstationarity are present.}
    \label{fig:sim1_setting4}
\end{figure}

\paragraph{Setting 4} 
The same stationary signals are used as above. The nonstationary signals, $\bo
n_1$, $\bo n_2$, and $\bo n_3$ correspond to those of Settings 1, 2, and 3,
respectively. We again use the ball kernel with radius $r = 3.4$.
\Cref{fig:sim1_setting4} shows that the results follow our expectations based on
the previous settings: \textsc{spSSAsir} finds one nonstationary signal,
\textsc{spSSAsave} and \textsc{spSSAcor} find two, and \textsc{spSSAcomb}
converges to zero. For this set of parameters, each individual method performs
the best on the 3-by-3 partition, but \textsc{spSSAcomb} converges the fastest
on the 4-by-4 partition. As long as there are enough points per part, increasing
granularity increases performance.

We have established that the proposed methods discriminate the two subspaces
successfully under the unrealistic assumption that the dimensions of the
subspaces are known. In the following, we suggest a procedure to estimate $q$.

\section{Estimating the nonstationary subspace dimension with
augmentation}\label{sec:rank_est}

So far, we have assumed that the subspace dimensions are known. However, in
practice this assumption is clearly unrealistic. It is worth noting that even in
the time series literature there is no generally established methodology for
estimating these dimensions.

In the following, we introduce an estimator for the
nonstationary subspace dimension. The estimator can be employed across all the
spSSA methods discussed in \Cref{sec:ssa}. This is possible because the they share a common underlying structure:

\begin{enumerate}
  \item Whiten the data to obtain $\xst$.
  \item Compute a matrix $\M(\xst)$, which is expected to have $q$ non-zero
  eigenvalues and $p-q$ zero eigenvalues. Since the signs of the eigenvalues do
  not matter for our purposes, we consider their absolute values and assume the
  ordering
  \begin{equation*}
    d_1 \geq \cdots \geq d_q > 0 = \cdots = 0.
  \end{equation*}
  \item Decompose the matrix as
  \begin{equation*}
    \M(\xst) = \V \D \V^\top,
  \end{equation*}
  where $\D$ is the diagonal matrix of eigenvalues and $\V$ contains
  the corresponding orthonormal eigenvectors.
\end{enumerate}
In the case of \Cref{sec:comb}, we get multiple $\M$ matrices, so the framework
does not apply directly. However, we do get a decomposition in terms of
pseudo-eigenvalues $\D$, and the pseudo-eigenvector matrix $\V$ which are the
necessary components for the following.

Based on the structure of $\M$, we estimate the nonstationary
subspace dimension $q$ with the so-called augmentation approach which was
originally proposed by \citet{LuoLi2020} in the context of dimension reduction
for i.i.d.\ vector-valued data. This framework has since been extended to
i.i.d.\ matrix- and tensor-valued observations
\citep{RadojicicLietzenNordhausenVirta2021,RadojicicLietzenNordhausenVirta2025},
as well as to time series data \citep{RadojicicNordhausen2024}. Nevertheless, to
the best of our knowledge, no corresponding extension to spatial data or to the
setting of stationary subspace analysis has been developed.

\paragraph{Augmentation procedure.}

Let $r \geq 1$ be fixed. Let $\e(\u) \in \mathbb{R}^r$ denote a white noise
random field, independent of $\xu$. This is to say that $\e(\u)$ is a
stationary random field satisfying Assumption (A1). The white noise random
field is used to construct the \emph{augmented standardized random field}
\begin{equation*}
  \xstu^* =
  \begin{pmatrix}
    \xstu \\
    \e(\u)
  \end{pmatrix}.
\end{equation*}

Under the assumed model, the associated matrix becomes
\begin{equation*}
  \M({\xst}^*) =\V^* \begin{pmatrix}
  \D & \bo 0 \\
  \bo 0 & \D_{\bs \varepsilon} 
  \end{pmatrix}\V^{* \top} =\V^* \begin{pmatrix}
  \D & \bo 0 \\
  \bo 0 & \bo 0
  \end{pmatrix}\V^{* \top},
\end{equation*}
where $\V^*$ is the diagonalizer of $\M({\xst}^*)$ and $\D_{\bs \varepsilon}$ is
a diagonal matrix corresponding to the eigenvalues associated with the white
noise. Since the white noise is stationary, we know that the associated
eigenvalues are zero. Furthermore, this tells us that
$\mathrm{rank}(\M({\xst}^*)) = \mathrm{rank}(\M(\xst)) = q$, i.e., the
augmentation does not affect the rank of $\M$.

Actually, we can say even more about the structure of the augmented matrices.
For each of the scatter matrices $\M_\mean$, $\M_\var$, $\M_{\cor, f}$, it holds
that 
\[
        \M({\xst}^*) = \begin{pmatrix}
          \M({\xst}) & \bo 0 \\
          \bo 0 & \bo 0
          \end{pmatrix}.
\]
This follows from the properties of scatter matrices and the
assumptions on the present signals. For example, in the case of $\M_\var$, we
can examine each summand separately. For each $\U_k$, the covariance matrix
$\Cov_{\U_k}({\xst}^*)$ is given by
\begin{equation}
  \label{eq:matrix}
        \E\left[\begin{pmatrix}
            \xstu - \E[\xstu ]\\
            \bs \varepsilon(\u) - \E[\bs \varepsilon(\u)]
        \end{pmatrix}\begin{pmatrix}
            \xstu^\top - \E[\xstu^\top ] \\
            \bs \varepsilon(\u)^\top - \E[\bs \varepsilon(\u)^\top]
        \end{pmatrix}^\top  \right] = \E\left[ \begin{pmatrix}
            \xstu \xstu^\top & \xstu \bs\varepsilon(\u)^\top \\
            \bs\varepsilon(\u)\xstu^\top & \bs\varepsilon(\u)\bs\varepsilon(\u)^\top
        \end{pmatrix} \right].
\end{equation}
Looking at the matrix on the right, we make the following observations: the top left block is simply $\Cov(\xst)$.
Next, since
$\bs\varepsilon(\u)$ is independent of $\xstu$, \[ \E[\xstu
\bs\varepsilon(\u)^\top] = \E[\xstu] \E[\bs\varepsilon(\u)^\top] = \bs 0 =
\E[\bs\varepsilon(\u)] \E[\xstu^\top] = \E[\bs\varepsilon(\u)\xstu^\top]. \]
Finally, $\E[\bs\varepsilon(\u)\bs\varepsilon(\u)^\top] = \Cov(\bs\varepsilon) =
\bo I_r$, and thus,
\[
    \bo I_{p+r} - \Cov({\xst}^*) = \bo I_{p+r} - \begin{pmatrix}
        \Cov(\xst) & \bo 0\\
        \bo 0 & \bo I_r
    \end{pmatrix} = \begin{pmatrix}
        \bo I_p - \Cov(\xst) & \bo 0\\
        \bo 0 & \bo 0
    \end{pmatrix}.
\]
Since we are only summing scaled squares ($A^2 = AA^\top $) of matrices of this
form, the resulting matrix, $\M_\var$ is of the same form. The arguments for the
other scatter matrices follows similarly.

Let us now take a closer look at the eigenvectors of $\M({\xst}^*)$, denoted by
\begin{equation*}
  \bo v_i^* = 
  \begin{pmatrix} 
    \bo v_i \\ 
    \bo v_i^{\aug} 
  \end{pmatrix},
\end{equation*}
where $\bo v_i \in \mathbb{R}^p$, $\bo v_i^{\aug} \in \mathbb{R}^r$, and $\bo
v_i^*$ corresponds to the eigenvalue $d_i^*$ of $\M({\xst}^*)$. Arrange the
eigenvalues such that
\begin{equation*}
  d_1^* \geq d_2^* \geq \cdots \geq d_{p+r}^*.
\end{equation*}
The eigenpair satisfies
\begin{equation*}
  \M({\xst}^*) \bo v_i^* = 
\begin{pmatrix}
          \M({\xst}) & \bo 0 \\
          \bo 0 & \bo 0
          \end{pmatrix}\begin{pmatrix} 
    \bo v_i \\ 
    \bo v_i^{\aug} 
  \end{pmatrix}
  =
  \begin{pmatrix} 
   \M(\xst) \bo v_i \\ 
   \bo 0
  \end{pmatrix} 
  = d_i^* \bo v_i^* 
  =
  \begin{pmatrix} 
   d_i^* \bo v_i \\ 
   d_i^* \bo v_i^{\aug}
  \end{pmatrix}.
\end{equation*}
This shows that $d_i^* \bo v_i^{\aug} = \bo 0$, for all $i$, and gives us
\begin{equation*}
  |d_i^*| \|\bo v_i^{\aug}\| = 0.
\end{equation*}

For all $i \leq q$, the eigenvalues $d_i^*$ are non-zero, and hence it follows
that the augmented part of the eigenvector has the norm $\|\bo v_i^{\aug}\| =
0$. However, this may not hold for $i > q$, since $d_i^* = 0$. Hence, the norms
of the augmented components $\bo v_i^{\aug}$ provide insight into the rank $q$
of $\M(\xst)$. Using the information provided by the eigenvectors and
eigenvalues, we present three methods for rank estimation. \\

\paragraph{Rank estimation via augmented eigenvectors.}

Let $\bo v_0^{\aug, j} := \bo 0$ and define the function
\begin{equation*}
  f:\{0,1,\dots,p\} \to \mathbb{R}, \quad f(i) = \frac{1}{s} \sum_{j=0}^s
  \|\bo v_i^{\aug, j}\|^2,
\end{equation*}
where $s$ is the number of repetitions of the augmentation procedure to average
out the randomness for simulating the white noise fields.

From the properties of the augmented parts of the eigenvectors, we know that
$f(i)$ should be approximately zero for $i \leq q$. Hence the point at which $f$
starts increasing gives us an estimate of the dimension.
\newline

\paragraph{Normalized scree plot.}

Define $d_{p+1} := 0$, and following \cite{LuoLi2020}, define the
normalized scree function by
\begin{equation*}
  \Phi:\{0,1,\dots,p\} \to \mathbb{R}, \quad \Phi(l) = \frac{d_{l+1}}
  {\sum_{i=1}^{l+1} d_i}.
\end{equation*}

The scree function starts at value $\Phi(0) = 1$, and goes down to zero once $i
> q$. So the sufficiently large non-zero value of the scree function gives an
estimate of $q$. Note that this estimate assumes that there is at least one
nonstationary signal. If all signals were to be stationary, all of the
eigenvalues $d_i$ would be zero and the values of $\Phi(\ell)$ would not be
defined for any $\ell$. The problem arises from the division by zero. This can
be easily mitigated by replacing the denominator by $1 + \sum_{i=1}^{l+1} d_i$.
However, it should be kept in mind that doing so breaks the scale invariance of
the estimate. For our purposes, we can ignore this issue as our model assumes by
default that there  is at least one nonstationary signal.  

\paragraph{Augmentation estimator.}

The final estimator combines both the eigenvalue and eigenvector information:
\begin{equation*}
  g:\{0,1,\dots,p\} \to \mathbb{R}, \quad g(k) = \Phi(k) + \sum_{i=1}^k f(i),
\end{equation*}
and the estimator of the number of signal components is given by
\begin{equation*}
  \hat{q} = \argmin_{k \in \{0,\dots,p\}} g(k).
\end{equation*}

In the ideal case, the scree function is non-zero and decreasing up to $q$. For
those values, $f$ is zero. This relation flips after $q$, at which point $f$
will be non-zero and increasing, and the scree function vanishes. Consequently,
the two functions intersect at $q$ producing a cusp in $g$. This point defines
the augmentation estimator. 

In practice, rank estimation based on augmented eigenvalues or normalized scree
plots typically requires graphical inspection, with the change point selected
visually. This subjective step may lead to ambiguity and a lack of
reproducibility. In contrast, plotting the augmentation estimator produces a
so-called ladle plot, which ideally exhibits a clear minimum and therefore
yields a unique and objective estimate of the rank. For this reason, we focus on
the augmentation estimator in the simulation studies presented below.

The three estimators introduced above are formulated for settings that can be
expressed within a generalized eigenvalue–eigenvector framework. When estimators
are combined, as in the case of \texttt{spSSAcomp}, the problem is instead cast
as one of joint diagonalization. Nevertheless, the augmentation procedure
remains applicable: in this setting, the eigenvalues are replaced by sums of the
absolute values of the corresponding pseudo-eigenvalues.

\section{Simulations II}\label{sec:sim2}

In the following simulations, we study the finite sample performance of the augmentation estimator.
We use Setting 4 from \Cref{sec:sim1} with the side length $\ell = 60$ and the
sample size $n = 3600$. 
We use a 4-by-4 grid partitioning of the domain. We test the method for
varying numbers of dimensions for the augmented data, namely, for $r = 1, 5, 10,
15$. For all dimensions of the augmented data, we perform the augmentation estimation procedure for 2000 repetitions. 
Note that the true nonstationary dimension is $q = 3$. 
We then compute how many times
each result appeared and measure the appearance rate of each computed value. The
results are presented as proportions of the horizontal bar with different values
being assigned different colors. The correct value is represented by the color
gray. Underestimations are represented by blue and overestimates by red. The
darker the color, the more we have over, or, underestimated. We fix the number
of trials $s$ for the augmented vectors to $s = 10$. This is a standard
assumption in the literature and is done for example in \cite{LuoLi2020,
RadojicicNordhausen2024}.

\begin{figure}[ht!]
    \centering
    \includegraphics[width=0.9\linewidth]{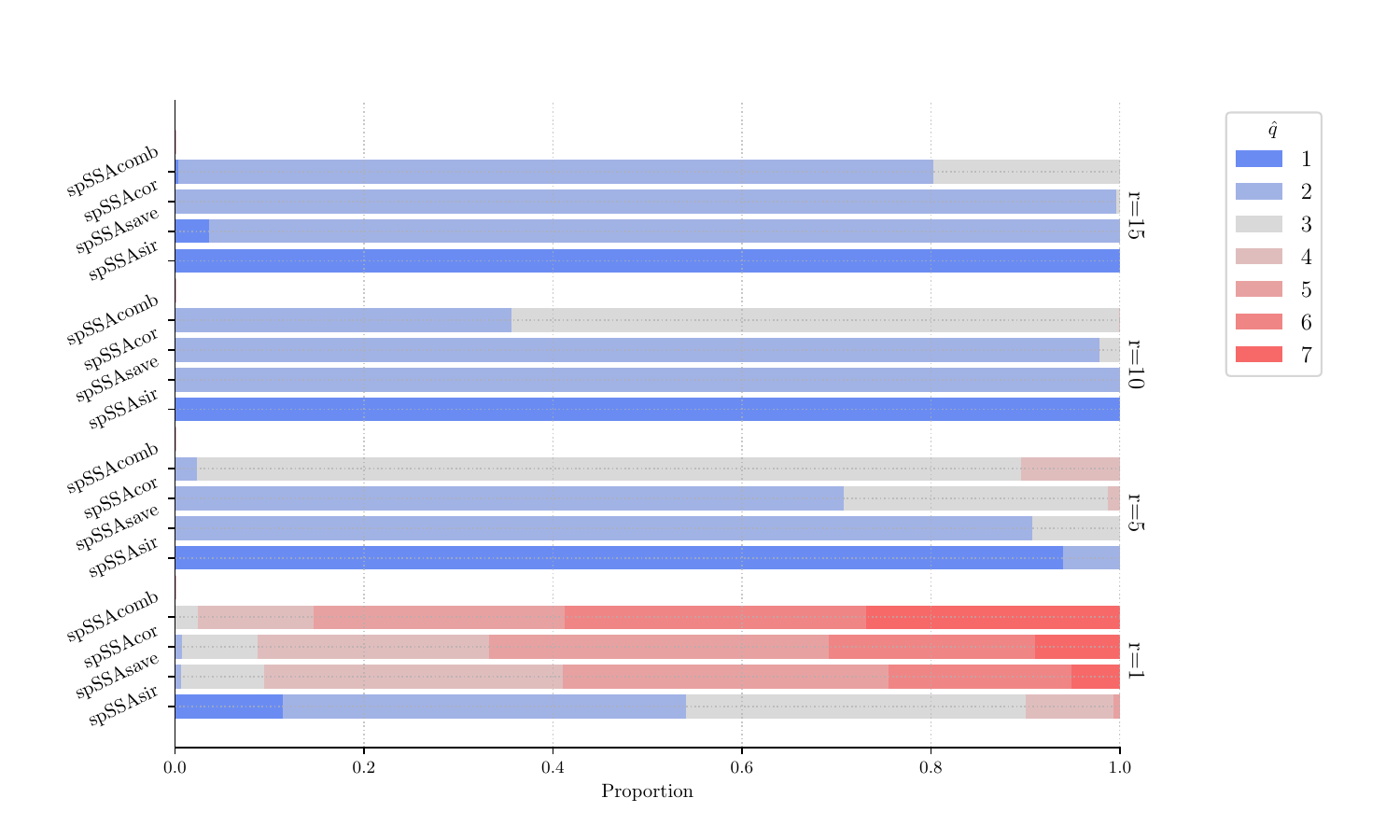}
    \caption{Rank estimation for Setting 4 using augmented noise dimensions $r=
    1, 5, 10, 15$. The simulation is done so that each signal has 3600 data
    points and the mixed signals are processed using a $(4, 4)$ partition of the
    domain. Neutral gray indicates successful estimation. Blue is used to
    indicate underestimation and red overestimation. The darker the color, the
    worse the estimation.}
    \label{fig:rank4}
\end{figure}

The results of the simulations can be seen in \Cref{fig:rank4}. For $r=1$, most
methods tend to overestimate. As we increase the number of augmented signals,
each method starts to underestimate. This happens because the ``mass'' of the
signals falls on the augmented part which is purely stationary. 

The results reflect what was expected: \textsc{spSSAsir} finds one signal, its
own, \textsc{spSSAsave} and \textsc{spSSAcor} find two signals, and the jointly
diagonalized estimate correctly estimates the nonstationary dimension. We see
that \textsc{spSSAcomb} gets the best results for $r= 5$ and $r = 10$. Thus, we
recommend $r$ be in the range $[5, 10]$ as a suitable choice of parameters. This
is in line with other recommendations in the literature, see for example
\cite{RadojicicVirta2025}, who show that for too large $r$, a high-dimensional
regime might be obtained which, if not properly handled, induces bias when
estimating the eigenvalues. \Cref{apx:add_plot} shows similar plots for the
different settings given in \Cref{sec:sim1}.

\section{Spatial stationary subspace analysis for the Kola moss data}\label{sec:kola}

In this section, we apply spSSA to the moss data from the Kola project and
estimate the dimension of the underlying stationary subspace. The data are
publicly available in the \texttt{R} package \texttt{StatDA} \cite{StatDA} and
are described in detail in \cite{reimann2011environmental}. The same data set
was also analyzed as a real-data example in
\cite{NordhausenOjaFilzmoserReimann2015} under the assumption that it is
stationary.

From the full data set out of 39 elements, we consider all
elements without missing values. Thus, for our analysis, we
include the elements Ag, Al, As, B, Ba, Be, Bi, Ca, Cd, Co, Cr, Cu, Fe, Hg, K,
La, Mg, Mn, Mo, Na, Ni, P, Pb, Rb, S, Sb, Sc, Se, Si, Sr, Th, Tl, U, V, Y, and
Zn, while excluding the elements Au, Pd, and Pt. Consequently, the resulting
random field $\xu$ is $36$-dimensional. The random field was measured in total
at 594 locations.

The observed data are compositional in nature and therefore require an
appropriate transformation before spSSA can be applied. Following the approach
of \cite{NordhausenOjaFilzmoserReimann2015}, to which we also refer for further
details, we consider $\x_{\mathrm{ilr}}$, the \emph{isometric log-ratio} (ilr)
transformation of the data. Since the ilr transformation maps a $D$-part
composition to $\mathbb{R}^{D-1}$, this transformation reduces the dimension by
one, yielding a $35$-dimensional random field.

\begin{figure}
    \centering
    \includegraphics[width=0.8\linewidth]{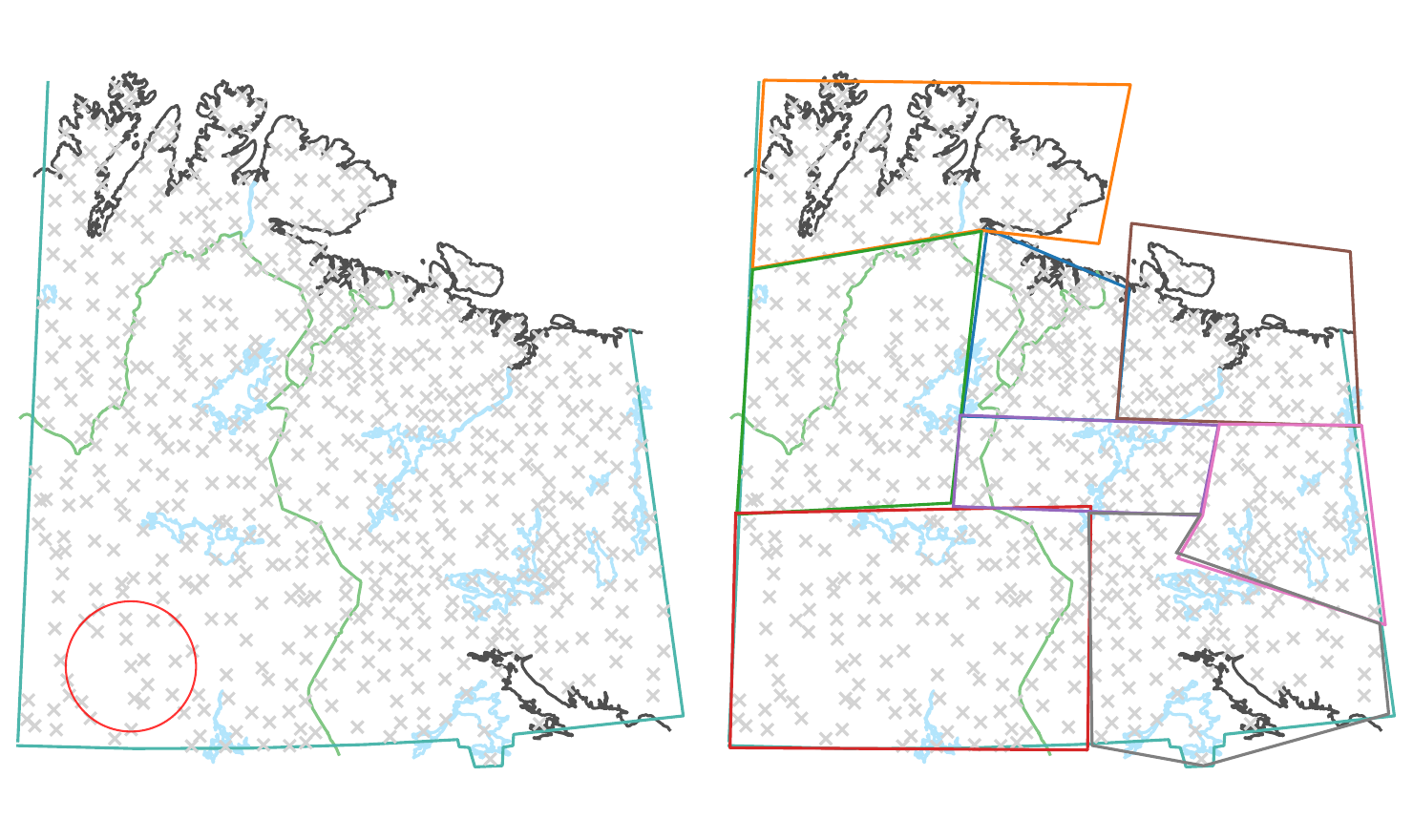}
    \caption{Two maps of the measurement locations for the Kola data. The measurement locations are marked with a gray $\times$ symbol. On the left, the red circle shows a ball of radius 50km. On the right, the used partition is shown.}
    \label{fig:kola_map}
\end{figure}

We apply our preferred method, \textsc{spSSAcomb}, which requires the
specification of a kernel function and a partition of the spatial domain. We
choose a ball kernel with radius $50$\,km, matching the choice in
\cite{NordhausenOjaFilzmoserReimann2015}. The partition of the domain is shown
in \Cref{fig:kola_map}. Note that we are again using the scaled version of
\textsc{spSSAcor} as a subroutine of \textsc{spSSAcomb}.

In \Cref{fig:kola_map}, the spatial domain is displayed twice. The left panel
shows the measurement locations together with an illustration of the ball kernel
with radius $50$\,km. The right panel displays the partition used in the
analysis, demonstrating that the proposed methodology is applicable also for
irregularly shaped partitions.

\begin{figure}[h!]
    \centering
    \includegraphics[width=0.8\linewidth]{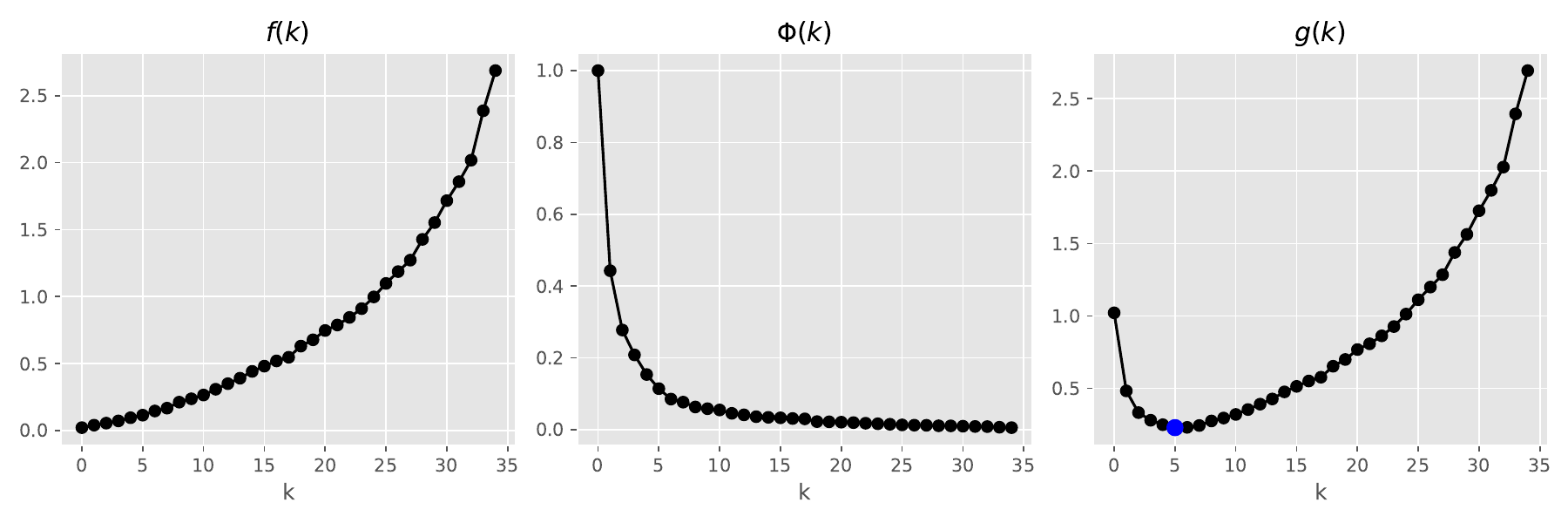}
    \caption{Plot of rank estimators $f, \Phi, g$ on the Kola data with the estimated rank $\hat{q} = 5$, highlighted in blue on the $g$ ladle.}
    \label{fig:ladle}
\end{figure}

We now estimate the dimension $\hat{q}$ of the nonstationary subspace of
$\x_{\mathrm{ilr}}$. To this end, we apply \textsc{spSSAcomb} using the kernel
function and the partition described above and set the tuning parameters to $r =
10$ and $s = 10$. The three criteria for rank estimation, $f(k)$, $\Phi(k)$, and
$g(k)$, are visualized  in \Cref{fig:ladle}. The minimum of $g(k)$ in the ladle
plot is highlighted in blue, providing the estimate $\hat{q} = 5$.

While the estimators $f(k)$ and $\Phi(k)$ individually do not indicate a clear
choice for $\hat{q}$, the combined criterion $g(k)$ exhibits a well-defined
minimum, illustrating for example the advantage of the ladle plot over a scree
plot.

\begin{figure}[!ht]
    \centering
    \includegraphics[width=\linewidth]{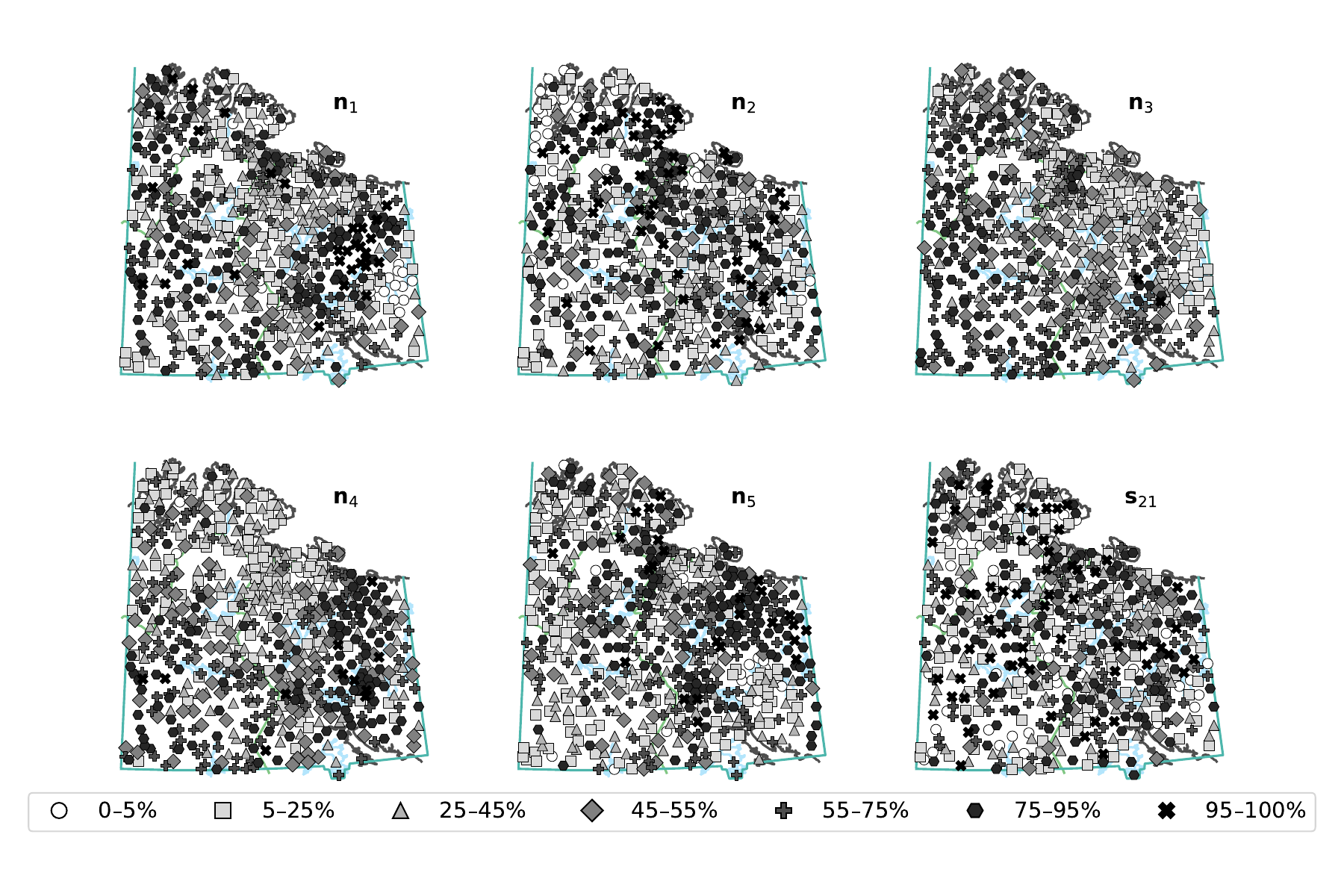}
    \caption{Nonstationary components and one stationary component of
    $\x_{\mathrm{ilr}}$. Symbols indicate percentile ranges of values.}
    \label{fig:kola_ssa}
\end{figure}

Based on the estimated rank $\hat{q} = 5$, the unmixing matrices $\W_{\comb, n}$
and $\W_{\comb, s}$ were constructed. Using these matrices, we separate the two
subspaces and visualize all five nonstationary signals as well as one randomly
chosen representative stationary signal $\bo s_{21}$. The resulting spatial
patterns are shown in \Cref{fig:kola_ssa}. In the figure, the values of the
extracted signals are displayed using symbols corresponding to percentile
ranges. We also give the pseudo-eigenvalues of the nonstationary signals and the
representative stationary signal for all $\M$ matrices involved in the joint
diagonalization in \Cref{tab:pseudoeigs}.

\begin{table}[!ht]
\centering
\caption{The pseudo-eigenvalues of the $\M$ matrices from approximate joint
diagonalization for the $\hat{q} = 5$ nonstationary signals and the
representative stationary signal $\bo s_{21}$. Note that since there are 5
nonstationary signals, the 21st stationary signal corresponds to the 26th
pseudo-eigenvalue.}
\label{tab:pseudoeigs}
\begin{tabular}{lcccccc}
\toprule
 
& $d_{1}$ 
& $d_{2}$ 
& $d_{3}$ 
& $d_{4}$ 
& $d_{5}$ 
& $d_{26}$ \\
\midrule
$\M_\mean$ & 0.025 & 0.139 & 0.766 & 0.622 & 0.054 & 0.107\\
$\M_\var$  & 3.251 & 1.899 & 0.869 & 0.960 & 1.639 & 0.486\\
$\M^s_\cor$  & 3.111 & 3.038 & 2.772 & 2.596 & 1.950 & 0.471\\
\bottomrule
\end{tabular}
\end{table}

The results indicate a clear contrast between the two components: the stationary
signal exhibits a fairly uniform spatial distribution across all value ranges,
whereas the nonstationary signals display pronounced regional structures and
spatial trends. The pseudo-eigenvalues of the individual matrices employed in
\textsc{spSSAcomb} provide additional insight into the nature of the
nonstationarity present in the extracted signals. We report these values for the
five nonstationary components in \Cref{tab:pseudoeigs}. An inspection of the
pseudo-eigenvalues suggests that the dominant sources of nonstationarity arise
from spatial dependence and spatially varying variance, with only a minor
contribution being attributable to spatial location trends.

\section{Conclusion}

In this paper, we developed tools for spatial stationary subspace analysis. We
showed via simulations that given the dimension of the nonstationary subspace,
we can successfully separate the two subspaces. If only a single type of
nonstationarity is present, the method developed specifically for such a case
performs best with \textsc{spSSAcomb} being a close second. If multiple types of
nonstationarities are present, \textsc{spSSAcomb} clearly achieves the best
performance. It is hence our recommendation to use \textsc{spSSAcomb}. The
importance of a proper partition was also demonstrated. It is recommended to use
a granular partition so that all parts still have enough data points to do
proper estimations. 

In any SSA method, the estimation of the subspace dimensions is of central
practical importance. In this paper, we propose three estimators: one based on
information contained in the eigenvectors, one relying on the eigenvalues, and a
third estimator that combines both sources of information. 

The combined estimator is closely related to the ladle estimator originally
introduced for i.i.d. data and stationary time series in
\citet{LuoLi2016,NordhausenVirta2018}. A key difference is that, in its original
formulation, the eigenvector-based component relies on bootstrapping. While
feasible in i.i.d. and time series settings, such resampling becomes
substantially more computationally demanding in a spatial framework. The
approach proposed here avoids this difficulty and is therefore better suited to
spatial data analysis. 

It is also worth noting that the stationary subspace analysis (SSA) methods for
time series discussed in \citet{FlumianMatilainenNordhausenTaskinen2024} share a
structural similarity with the spSSA methods considered in this paper.
Consequently, the dimension estimation strategy developed here can be directly
transferred to these SSA methodologies with minimal adaptation.

The second simulation study demonstrates that the augmentation
estimator—combining information from both eigenvalues and
eigenvectors—successfully recovers the dimension of the nonstationary subspace.
In addition, the results consistently indicate that \textsc{spSSAcomb} provides
the most reliable performance across scenarios. Our recommendations for tuning
parameter choices are consistent with existing literature: specifically, we
suggest fixing $s = 10$ and selecting $r$ from the range $[5, 10]$.

All in all, the proposed methodology equips practitioners with a practical and
robust set of tools for separating multivariate spatial data into stationary and
nonstationary components. These components can subsequently be modeled
independently using appropriate techniques, which is typically far more
tractable than analyzing the original high-dimensional data in which both
structures are confounded. This practical value is illustrated through the
analysis of the Kola moss data set, where a 35-dimensional random field is
effectively decomposed into a five-dimensional nonstationary component and a
30-dimensional stationary component.

\section*{Acknowledgments}

We acknowledge computational resources from CSC -- IT Center for Science,
Finland. PS and KN were supported by the Research Council of Finland (363261).
AMR acknowledges funding from the French National Research Agency (ANR) under
the Investissements d’Avenir program (ANR-17-EURE-0010).

\appendix

\section{Comparing \textsc{spSSAcor} to scaled \textsc{spSSAcor}}\label{apx:comp}

\begin{figure}[!ht]
    \centering
    \includegraphics[width=0.8\linewidth]{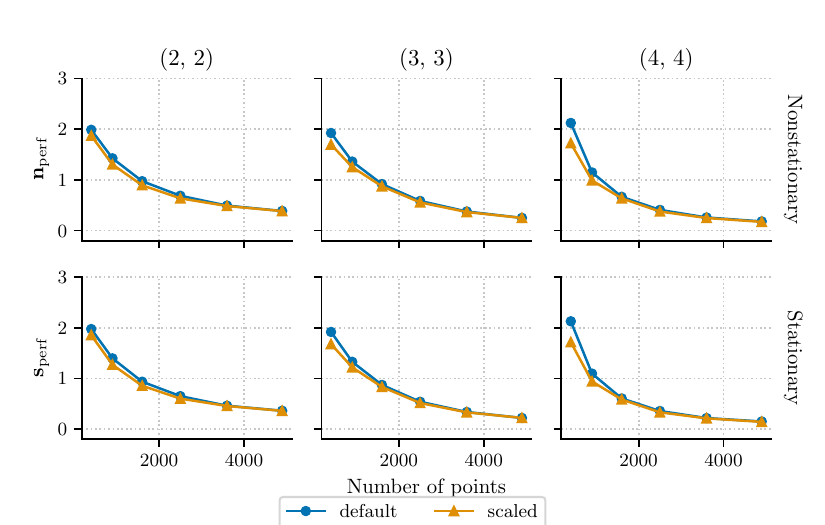}
    \caption{Comparison between \textsc{spSSAcor} (default) and scaled
    \textsc{spSSAcor} (scaled) on data from Setting 3 from \Cref{sec:sim1}.}
    \label{fig:lcor_comp}
\end{figure}

To demonstrate the comparability in terms of performance for the
\textsc{spSSAcor} methods, we repeat the simulations of Setting 3 from
\Cref{sec:sim1}. This time, we only include two curves in the resulting plots:
one for the unscaled \textsc{spSSAcor}, denoted default, and on for the scaled
version, denoted scaled. The results can be seen in \Cref{fig:lcor_comp}. The
results demonstrate clearly that the scaled \textsc{spSSAcor} causes no loss in
performance, and in fact performs slightly better than its unscaled counterpart. 

Let us also demonstrate the need to use the scaled version by comparing norms of
the $\M$ matrices. Consider the scenario where the 5-variate stationary signal
$\s$ from \Cref{sec:sim1} on 2500 data points is mixed with some orthogonal
matrix $\A \in \R^{5 \times 5}$ to get $\xu = \A\s$. Consider further the
standardized data $\xstu$. Let us now compute $\M_\mean$, $\M_\var$, $\M_{\cor,
f}$ and $\M^s_{\cor, f}$ from $\xstu$. For the kernel function $f$, we use the
ball kernel with radius $2.2$. In \Cref{tab:norm_comp}, we have tabulated the
average norm of these $\M$ matrices over 1000 repetitions using three different
$k$-by-$k$ partitions for $k = 2, 3, 4$. This data shows that the norms of all
$\M$ matrices increase as the number of partitions increases, but the $\M_{\cor,
f}$ matrices differ by two orders of magnitude from the other $\M$ matrices.
Even though $\M^s_{\cor, f}$ is still of greater magnitude than $\M_\mean$ and
$\M_\var$, it is at least of comparable scale. Note that we chose to demonstrate
this effect on stationary signals since the mean and variance estimators are
close to zero, making the need for scaling more apparent. 

\begin{table}[!ht]
\centering
\caption{Average norms of $\M$ matrices for a mixed stationary 5-variate standardized random field.}
\label{tab:norm_comp}
\begin{tabular}{ccccc}
\toprule
Partition 
& $\M_\mean$ 
& $\M_\var$ 
& $\M^{s}_{\cor,f}$ 
& $\M_{\cor,f}$ \\
\midrule
$(2,2)$ & 0.028 & 0.044 & 0.084 &  5.364 \\
$(3,3)$ & 0.056 & 0.105 & 0.193 & 11.727 \\
$(4,4)$ & 0.092 & 0.181 & 0.323 & 18.274 \\
\bottomrule
\end{tabular}
\end{table}
\clearpage

\begin{figure}[!ht]
    \centering
    \includegraphics[width=0.8\linewidth]{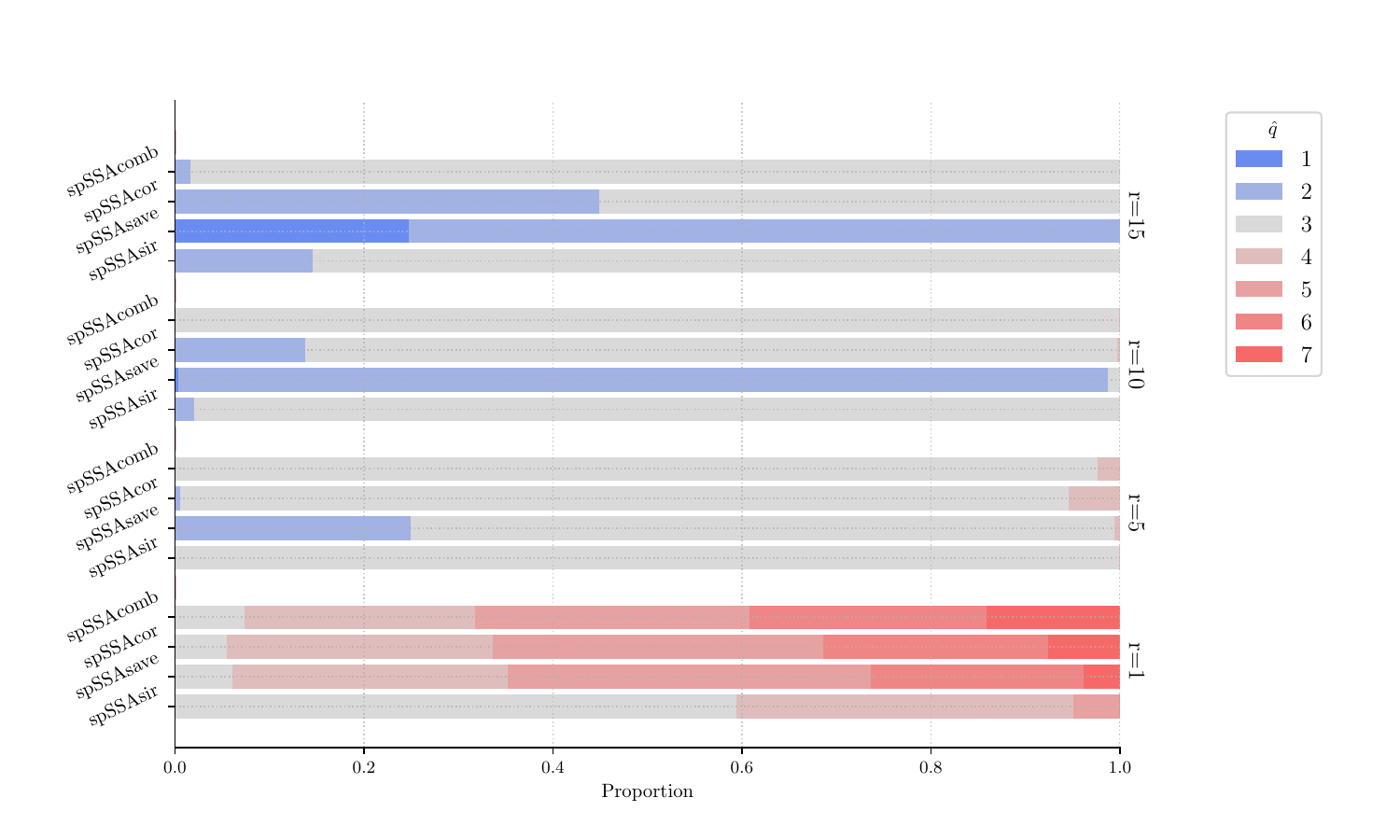}
    \caption{Rank estimation for Setting 1 using augmented noise dimensions $r=
    1, 5, 10, 15$. The simulation is done so that each signal has 3600 data
    points and the mixed signals are processed using a $(4, 4)$ partition of the
    domain. Neutral gray indicates successful estimation. Blue is used to
    indicate underestimation and red overestimation. The darker the color, the
    worse the estimation.}
    \label{fig:rank1}
\end{figure}

\clearpage
\section{Additional simulations for the augmentation estimator}\label{apx:add_plot}

\begin{figure}[ht!]
    \centering
    \includegraphics[width=0.78\linewidth]{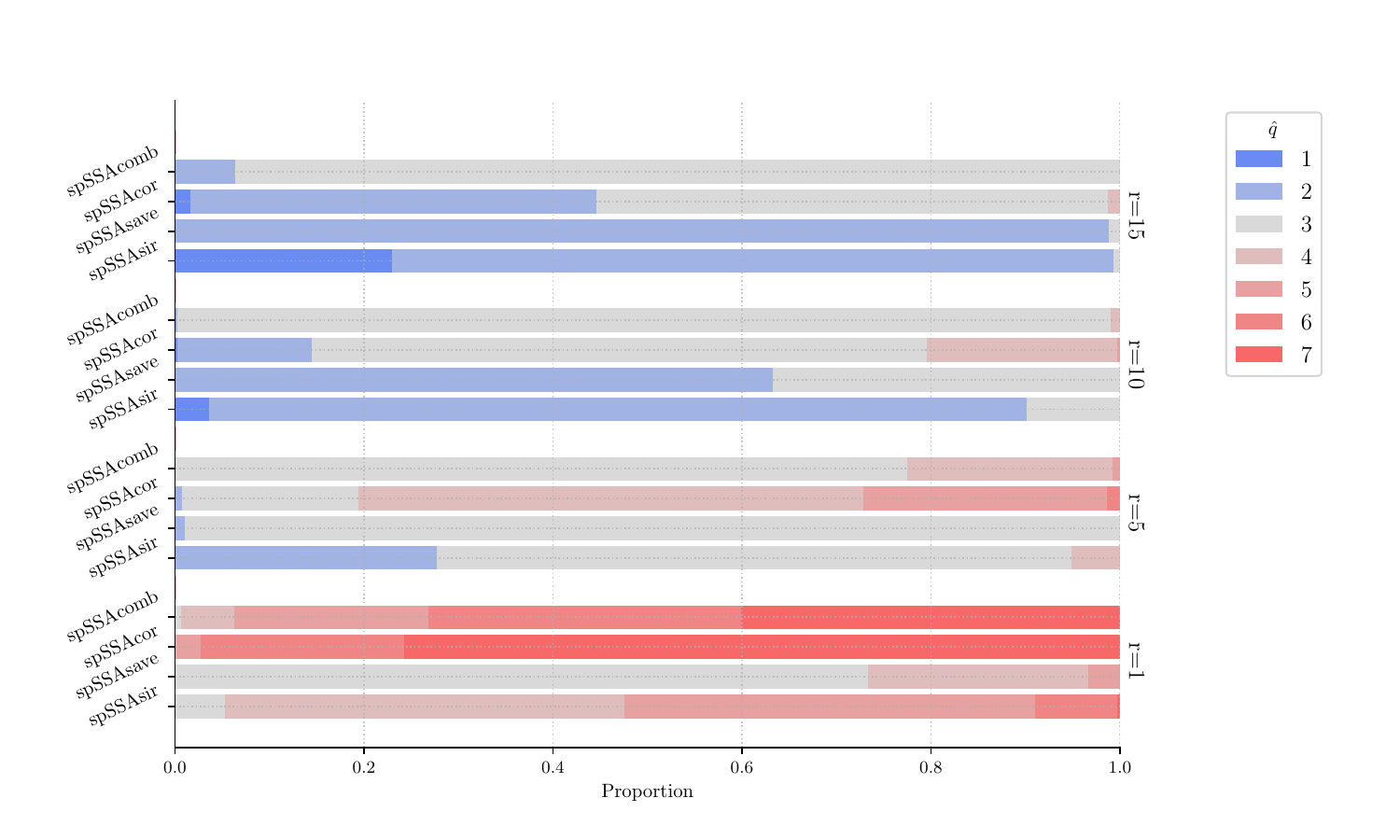}
    \caption{Rank estimation for Setting 2 using augmented noise dimensions $r=
    1, 5, 10, 15$. The simulation is done so that each signal has 3600 data
    points and the mixed signals are processed using a $(4, 4)$ partition of the
    domain. Neutral gray indicates successful estimation. Blue is used to
    indicate underestimation and red overestimation. The darker the color, the
    worse the estimation.}
    \label{fig:rank2}
\end{figure}

\begin{figure}[ht!]
    \centering
    \includegraphics[width=0.78\linewidth]{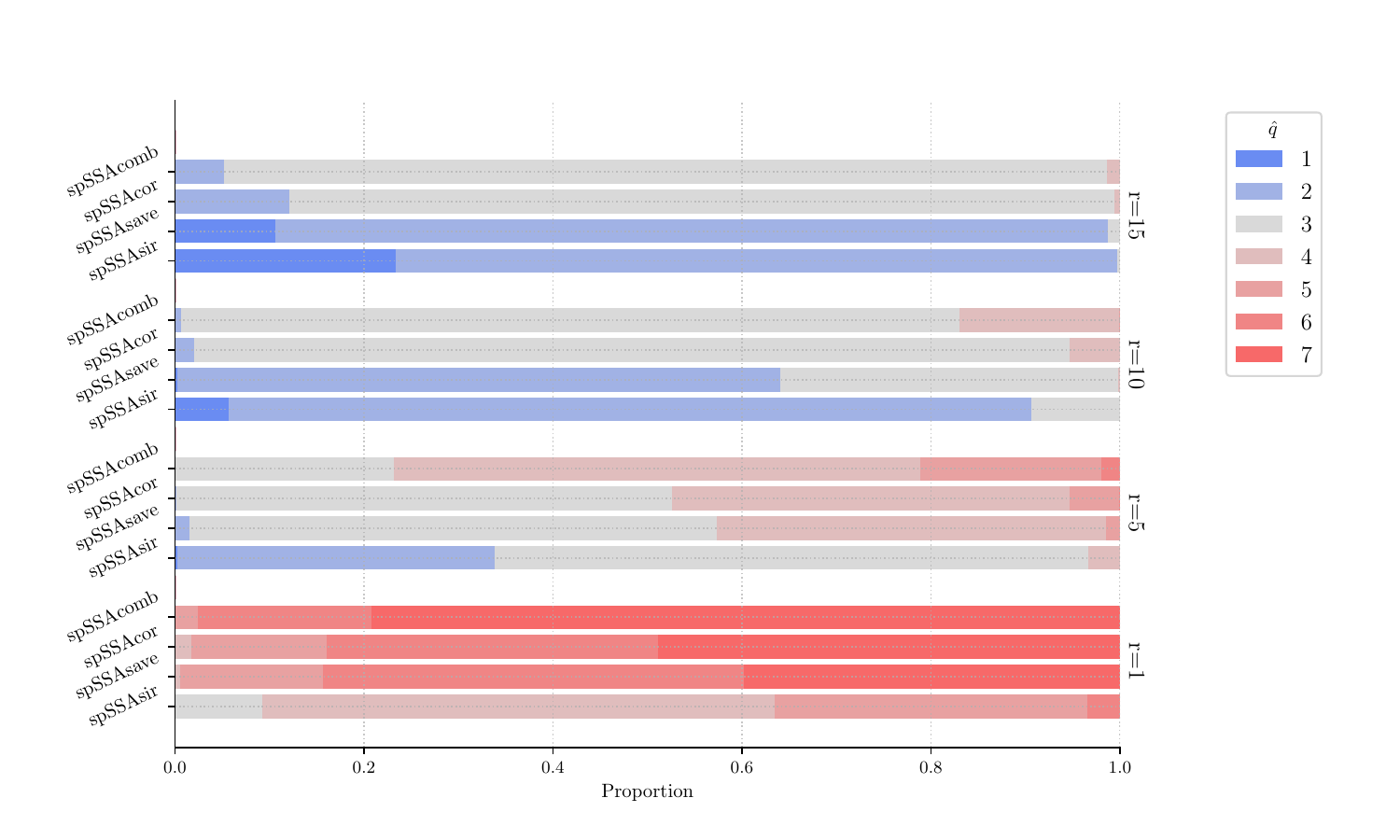}
    \caption{Rank estimation for Setting 3 using augmented noise dimensions $r=
    1, 5, 10, 15$. The simulation is done so that each signal has 3600 data
    points and the mixed signals are processed using a $(4, 4)$ partition of the
    domain. Neutral gray indicates successful estimation. Blue is used to
    indicate underestimation and red overestimation. The darker the color, the
    worse the estimation.}
    \label{fig:rank3}
\end{figure}

In this section, we repeat the simulation from \Cref{sec:sim2} for Settings 1,
 2, and 3 from \Cref{sec:sim1}. Everything is kept the same except the types of nonstationary signals.
 The results for these three settings can be seen in \Cref{fig:rank1},
 \Cref{fig:rank2}, and \Cref{fig:rank3}, respectively. The results show that if
 only one type of nonstationarity is present, we can determine the order even
 more accurately. We further see that the recommended values of $r$ work in all
 settings.

\bibliographystyle{apalike}  
\bibliography{references}  
\end{document}